\documentclass[fleqn,usenatbib]{mnras}
\usepackage{newtxtext,newtxmath}
\usepackage[T1]{fontenc}

\DeclareRobustCommand{\VAN}[3]{#2}
\let\VANthebibliography\thebibliography
\def\thebibliography{\DeclareRobustCommand{\VAN}[3]{##3}\VANthebibliography}
\usepackage{newtxtext,newtxmath}
 
\usepackage{graphicx}	
\usepackage{amsmath}	
\usepackage{subcaption}
\usepackage[dvipsnames]{xcolor}
\usepackage{newtxtext, newtxmath}
\usepackage{physics}

\usepackage[T1]{fontenc}
\usepackage{ae,aecompl}

\usepackage{graphicx}	
\usepackage{amsmath}	
\usepackage{bm}

\DeclareMathOperator*{\argmin}{arg\,min}

\usepackage{physics}

\def\btheta{\boldsymbol{\theta}}

\usepackage{orcidlink}

\title[Graph Neural Network for IA]{Galaxies and Halos on Graph Neural Networks: Deep Generative Modeling Scalar and Vector Quantities for Intrinsic Alignment}

\author[Jagvaral et al.]{
Yesukhei Jagvaral$^{1,2}$\thanks{E-mail: yjagvara@andrew.cmu.edu}\orcidlink{https://orcid.org/0000-0001-7068-7037},
Fran\c{c}ois Lanusse$^{3}$\orcidlink{0000-0001-7956-0542},
Sukhdeep Singh$^{1,2}$,
Rachel Mandelbaum$^{1,2}$\orcidlink{0000-0003-2271-1527},
\newauthor
Siamak Ravanbakhsh,$^{4,5}$ 
Duncan Campbell$^{1,6}$
\\
$^{1}$McWilliams Center for Cosmology, Department of Physics, Carnegie Mellon University, Pittsburgh, PA 15213, USA\\
$^{2}$NSF AI Planning Institute for Data-Driven Discovery in Physics, Carnegie Mellon University, Pittsburgh, PA 15213, USA\\
$^{3}$AIM, CEA, CNRS, Universit\'e Paris-Saclay, Universit\'e Paris Diderot, Sorbonne Paris Cit\'e, F-91191 Gif-sur-Yvette, France\\
$^{4}$School of Computer Science, McGill
University, Montreal, QC H3A 0G4, Canada\\
$^{5}$ Mila, Quebec AI Institute\\
$^{6}$Epistemix Inc., Pittsburgh\\
}

\date{Accepted XXX. Received YYY; in original form ZZZ}

\pubyear{2020}

\begin{document}
\label{firstpage}
\pagerange{\pageref{firstpage}--\pageref{lastpage}}
\maketitle

\begin{abstract}In order to prepare for the upcoming wide-field cosmological surveys,  large simulations of the Universe with realistic galaxy populations are required. In particular, the tendency of galaxies to naturally align towards overdensities, an effect called intrinsic alignments (IA), can be a major source of systematics in the weak lensing analysis. As the details of galaxy formation and evolution relevant to IA cannot be simulated in practice on such volumes, we propose as an alternative a Deep Generative Model. This model is trained on the IllustrisTNG-100 simulation and is capable of sampling the orientations of a population of galaxies so as to recover the correct alignments. In our approach, we model the cosmic web as a set of graphs, where the graphs are constructed for each halo, 
and galaxy orientations as a signal on those graphs. The generative model is implemented on a Generative Adversarial Network architecture and uses specifically designed Graph-Convolutional Networks sensitive to the relative 3D positions of the vertices. 
Given (sub)halo masses and tidal fields, the model is able to learn and predict  scalar features such as galaxy shapes; and more importantly, vector features such as the 3D orientation of the major axis of the ellipsoid and the complex 2D ellipticities. For correlations of 3D orientations the model is in good quantitative agreement with the measured values from the simulation,  except for at very small and transition scales. 
For correlations of 2D ellipticities, the model is in good quantitative agreement with the measured values from the simulation.
\end{abstract}


 \section{Introduction}

Upcoming wide-field cosmological surveys such as the Vera C.\ Rubin Observatory Legacy
Survey of Space and Time (LSST)\footnote{ \url{https://www.lsst.org/} },  Roman Space Telescope\footnote{ \url{https://roman.gsfc.nasa.gov/} } High Latitude Survey (HLS) and Euclid\footnote{ \url{https://www.euclid-ec.org/} }  will provide data that can be used to  answer fundamental questions on the 
nature of dark energy  through a precise measurement of cosmic shear, the observable correlations of galaxy shapes due to the minute but coherent deformations of distant
galaxy images by the gravitational influence of massive structures along the line of sight \citep[for a review, see][]{Kilbinger}. One major astrophysical contaminant
that arises when trying to measure this signal comes from the tendency of galaxies to naturally align with large-scale structure, an effect 
called  intrinsic alignments (IA) that  causes coherent shape distortions that anti-correlate with lensing shape distortions, and therefore bias the cosmological analysis if not accounted for  \citep{troxel-ishak}. 
High resolution
hydrodynamical simulations, which can simulate the formation and evolution of individual galaxies, are valuable tools 
to study these alignments,    
but remain limited to small cosmological volumes due to their computational costs. In this work,
we develop a deep generative model of 2D and 3D galaxy orientations that can capture the correct correlations of galaxy alignments, with a future goal of  using this model to inpaint realistic galaxy alignments in much larger N-body simulations at very little cost.

In order to extract robust cosmological information from weak lensing surveys, accurate and precise models of survey data are needed. Cosmological simulations are often employed to produce synthetic survey data (mock catalogs) to facilitate the design and provide a test of analysis pipelines and systematics mitigation methods. Cosmological simulations have seen rapid advances in recent decades. Dark matter only (DM-only) N-body simulations such as the Outer Rim  simulation achieved $\sim (4 \, \text{Gpc})^3$ scales with unprecedented resolutions exceeding 1 trillion particles \citep{outer-rim}.  For DM-only simulations, galaxies
must be included in post-processing; there are a variety of methods for doing so.  One method, semi-analytical models (SAMs) of galaxy formation and evolution, contain a number of free tunable parameters and can produce a synthetic galaxy population along with a variety of descriptive data \citep{somerville-SAM,guo-SAM}.
Even though SAMs have been successful in reproducing some observed galaxy properties \citep{somerville-dave}, in the coming era of precision cosmology their  simple  nature and challenges in matching galaxy populations across a range of redshifts 
leaves some uncertainty in the synthetic galaxy components of mock catalogs.

Another natural way to incorporate synthetic galaxies in mock catalogs is to directly implement them together with the dark matter with a hydrodynamic prescription and try to capture the full physics. In the past decade, cosmological hydrodynamical simulations 
such as MassiveBlack-II \citep{Khandai-mb2}, Illustris
\citep{Vogelsberger-illustris}, EAGLE \citep{Schaye-EAGLE}, Horizon-AGN \citep{dubois-horizon} and IllustrisTNG \citep{nelson-tng-publicdata}  have had some success  in producing realistic galaxies with properties that match those of observed galaxies to some degree. 
IA has been studied within these simulations \citep[e.g.,][]{tenneti-ia, eagle-ia,tenneti-disc-ellip,chisari-horizon-ia,samuroff-2020} in order to accurately model and constrain IA models. 
Still, these types of simulations are based on resolution elements  and along with the hydrodynamic equations, various astrophysical processes (such as AGN and stellar feedback; gas cooling; star formation) are based on effective sub-resolution models. Despite recent progress in computational astrophysics and cosmology, an {\em ab initio}  cosmological simulation  is far beyond current capabilities \citep{Vogelsberger-review}. 
However, as survey instruments become larger and more powerful, cosmological simulations have to encompass large volumes and high-resolution to keep up with the observational data.  Currently, hydrodynamical simulations cannot reach the desired Gpc scale and resolution scale for upcoming surveys.  Thus the best option to produce fast robust mock catalogs is to combine N-body simulations and some form of galaxy model, ideally non-parametric, as done for LSST DESC in \cite{cosmodc2}  and for Euclid\footnote{https://sci.esa.int/web/euclid/-/59348-euclid-flagship-mock-galaxy-catalogue} (based on \citealt{PKDGRAV3}).

It is now well established that galaxies form in DM halos, and thus galaxy evolution is inevitably tied to the growth and evolution of each galaxy's parent DM halo. In the literature, this inter-connectedness of galaxy and halos is dubbed the \textit{galaxy-halo connection} and is typically modeled as a multivariate distribution of various galaxy and halo properties. There has been some success in predicting and modeling  properties such as mass, abundance, clustering, through the use of the halo occupation distribution (HOD) and  subhalo abundance matching (SHAM)\citep{somerville-dave}. However,  there is mounting evidence that  the full galaxy-halo connection is highly non-linear and high dimensional  \citep[see, e.g.,][for a comprehensive review]{wechsler-tinker}. For example, when modeling the above-mentioned IA, the orientation of a galaxy is a 3D vector property (though we only observe the 2D projection of the shape) that correlates with its environment and the underlying large scale structure.  The halo models of IA were developed to capture the small scale orientations within the parent DM halo \citep{schneider-bridle,fortuna}. For large scale alignments, the linear alignment model \citep{catelan, hirata-seljak} and the later extensions  that included non-linear contributions \citep{bridle-king,blazek-tatt}, were successful in capturing the large-scale alignment of elliptical galaxies, despite underestimating the alignment at intermediate and small scales. For example, the tidal alignment-tidal torquing model from \cite{blazek-tatt} showed some promise for describing alignments down to $\sim$1 Mpc/h scales in \citet{samuroff-2020}; with this in mind, we would like our model to work to even smaller scales. 
However, these analytic models  include a number of tunable parameters that are challenging to physically interpret and rely on assumptions that may not be robust.

 In recent years, many fields of science are undergoing an Artificial Intelligence (AI) revolution and many AI models have been used in astrophysical and cosmological frameworks (see \url{github.com/georgestein/ml-in-cosmology} for a list and \citealt{ntampaka-2019} for the role of ML in cosmology). In particular, unsupervised learning methods are designed to learn and detect patterns from the data itself,  whereas  traditional numeric and semi-analytic approaches are based on physical laws and models that are known beforehand. One class of unsupervised deep learning methods are deep generative models (DGM), where the DGM is trained on a given dataset and learns the likelihood (explicitly or implicitly) that can be used to generate new sample data. In many cases, DGMs have been shown to outperform traditional numeric and semi-analytic models in accuracy or speed \citep{kodi-a, yin-li}.
 High-resolution hydro-sims mentioned above show promise for training and testing DGMs, given that we have access to the full 3D phase space data and numerous scalar features associated with galaxies and DM halos. Thus, in order to enable fast  production of mock catalog for future surveys, in this work we will train a DGM on hydrodynamical simulations, to capture the relevant scalar and vector features.
 
 In this work, we aim to capture the complex relation of the density field, DM halo and the galaxy with a deep generative model, and then sample from this model various scalar and vector features of halos and galaxies. 
Our approach is to model the cosmic web as a set of \textit{graphs}, where the graphs are constructed for each halo.  Graphs are a natural data structure to capture the correlations of galaxy properties
amongst neighbours \citep{gnn-review}, given that galaxies are distributed sparsely in the Universe. Graphs are defined as sets of \textit{vertices} (also called \textit{nodes}) and each of the \textit{edges}   (or \textit{links}) connecting pairs of vertices.  Given this structure, we adapt a Graph-Convolutional Network \citep{Defferrard2016, kipf-welling} to be sensitive
to the 3D relative positions between vertices, a key ingredient to make the model aware of the  Euclidean geometry of the problem beyond the graph connectivity. Using these layers, we subsequently implement a deep Generative Adversarial Network (GAN,  \citealt{Goodfellow-GAN}) for signals on graphs. 
As part of this work, we will explore different network architectures and different physical information content from the simulations, and identify which ones enable robust reconstruction of the large-scale galaxy alignments.

This paper is organized as follows: We begin in  \S \ref{DL_intro} by describing the deep learning methods that were used, such as our graph convolutional neural network construction and the GAN architecture.   Next, we begin \S \ref{methods-section}  by   describing the simulation suite we have used in \S \ref{simulation}, after which we introduce our estimators of the tidal field, galaxy shape and two-point statistics. Next, in \S \ref{neural_model} we describe the models in detail. 
The results are presented in \S \ref{results} and we conclude in \S \ref{conc}. 
  
\section{Deep Learning Background}\label{DL_intro}
In this section, we will introduce the graph convolutional networks we have implemented and then give a brief introduction to the generative adversarial network  architecture. Our end goal is to model a conditional probability density $p(\bm{y} | \bm{x})$ where $\bm{y}$ is the desired property of a galaxy that we want to model (for example, shape or orientation), and $\bm{x}$ are the features used as input to the model, such as the (sub)halo mass, position or the tidal field.

\subsection{Graph Convolutional Networks}

\subsubsection{Spectral graph convolutions}
 In this work, we are considering undirected and connected graphs, which can be defined as  $\mathcal{G} = (\mathcal{V} , \mathcal{E}, \mathbf{W})$, where $\mathcal{V}$ is the set of graphs vertices, with $\left\vert \mathcal{V} \right\vert = n$ the number of vertices,  $\mathcal{E}$ is the set of graph edges and $\mathbf{W} \in \mathbb{R}^{n \times  n}$ is the weighted adjacency matrix.
The normalized combinatorial graph  Laplacian is defined as $\mathbf{L} = \mathbf{I} - \mathbf{D}^{-1/2} \, \mathbf{W} \, \mathbf{D}^{-1/2}$, where $\mathbf{D} \in \mathbb{R}^{n \times  n}$ is the diagonal degree matrix with $D_{ii} = \sum_j W_{ij}$. 
Note that this operator is positive semi-definite and therefore  admits an eigenvalue decomposition defined as:
\begin{equation}
	\mathbf{L} = \mathbf{U} \mathbf{\Lambda} \mathbf{U}^t
\end{equation}
where $\mathbf{U}$ is a unitary matrix and $\mathbf{\Lambda} = \text{diag}([\lambda_0, \lambda_1, \cdots, \lambda_n])$ are the eigenvalues  of the operator. By analogy with a traditional Euclidean Laplacian, this transform is called a graph Fourier transform. The columns of $\mathbf{U}$ are called graph Fourier modes  and $\mathbf{\Lambda}$ is the diagonal matrix of graph Fourier frequencies. For a given signal $f \in \mathbb{R}^n$, the graph Fourier transform of $f$ is then defined as $\hat{f} = \mathbf{U}^t f$. 

Given this  harmonic transform, it becomes possible to define spectral filtering on graphs by defining a convolution product on graphs as a multiplication in Fourier space:
\begin{equation}
	f  \star g = \mathbf{U} \left( ( \mathbf{U}^t f) \odot  ( \mathbf{U}^t g ) \right) = \mathbf{U} \left(  \hat{f} \odot \hat{g} \right)
\end{equation}
where $\odot$ is the Hadamard product. While this expression allows for convolution operations  on graphs, it is a costly operation, as it first requires a decomposition of the  graph Laplacian as well as dense matrix vector multiplications. However, it can be approximated to first order in terms of the graph Laplacian as such (details in Appendix \ref{appendix:graph_appendix}): 
\begin{equation}
	g_\theta \simeq \mathbf{\theta}_0 +  2 \mathbf{\theta}_1 \mathbf{L}
	\label{eq:approx_conv}
\end{equation}

Then, we can use this first order approximation to parameterize graph convolutions, which can be implemented extremely efficiently by 
sparse matrix vector multiplication. We define one Graph Convolutional Network layer with an activation $y_i$ for a node $i$  as:
\begin{equation}
	\forall i \in \mathcal{V}, \	   y_i = \mathbf{b} + \mathbf{W}_0 h_i + \sum_{j \in \mathcal{N}_i} w_{i, j} \mathbf{W}_1 h_j
\end{equation}
where  $\mathbf{b} $  represents a vector of bias terms, we denote by $\mathcal{N}_i$ the set of immediate neighbors\footnote{Immediate neighbors or first neighbors are neighbors that are one hop away from node $i$.} of vertex $i$, $\mathbf{W}_0$ are the weights that apply a linear transform to the activation vector $h_i$ of node $i$ (i.e., self connection), $w_{i, j}$ are linear transforms on the activation vectors $h_j$ of the nodes $j$ in the neighborhood of $i$, and $\mathbf{W}_1$ are the set of weights that apply to the immediate neighbors. 
While the expressivity of a single GCN layer is quite limited,
by stacking a large number of them, complex mappings on graphs can be represented.

\subsubsection{Directional convolution kernels}

The GCN introduced above uses the same isotropic convolution kernels for the entire graph. However, given the  nature of our signal, we expect the 3D positions of neighbouring galaxies to be relevant to their alignments, for instance we know that within a halo, satellites tend to align towards the central galaxy \citep{radial-align}. 
We therefore want to design graph convolutions that have some sensitivity to 3D orientations. 

Based on the dynamic convolution kernels introduced in \cite{Verma2017}, we propose the following direction-dependent graph convolution layer:
\begin{equation}
	y_i=  \mathbf{b} + \mathbf{W}_0 h_i +  \sum\limits_{m=1}^{M} \frac{1}{|\mathcal{N}_i|} \sum_{j \in \mathcal{N}_i} q_m(\mathbf{r}_i, \mathbf{r}_j) \mathbf{W}_m h_j 
\end{equation}
here $|\mathcal{N}_i|$ denotes the cardinality of the set $\mathcal{N}_i$, $M$ is the number of directions, and $\mathbf{r}$ are the 3D Cartesian coordinates of the node. 
The  $q_m(\mathbf{r}_i, \mathbf{r}_j)$ 
are normalized so that $\sum_{m=1}^{M} q_m(\mathbf{r}_i, \mathbf{r}_j) = 1$ and are defined as:
\begin{equation}
		q_m(\mathbf{r}_i, \mathbf{r}_j) \propto \exp(- \mathbf{d}_m^t \cdot (\mathbf{r}_j - \mathbf{r}_i)) \  g_\lambda( \parallel \mathbf{r}_i - \mathbf{r}_j \parallel_2^2),
\end{equation}
where the $\{ \mathbf{d}_m \}_{m \in [1,M]}$ 
are a set of directions we want to make the kernel sensitive to, and $g_\lambda$ is a parametric function of the distance  between two vertices. With this parameterization, how a given vertex $i$ receives contributions from its neighbors will be a function of the directions to the neighbors, as well as a function of the distance. In this work we chose an exponential parametrization of the form: $g_\lambda(r) = \exp( - r^2/2\lambda^2)$, where $\lambda$ is fit automatically during training.

 Finally, when modeling 3D alignments, we use a set of 26 directions on the unit sphere, equivalent to the 27 voxels of a 3D 3x3x3 convolution kernel, minus the central voxel which in our graph neural network will correspond to node self-connections\footnote{In simpler terms, for a cube consisting of 3x3x3 voxels, the center voxel will have $3^3  -1=26$ connections; we subtract 1 because it accounts for the self-connection }.

\subsection{Fitting Implicit Distributions}

The other aspect of the problem is learning how to model, and sample from, a conditional probability density $p(\bm{y} | \bm{x})$ where $\bm{y}$ might be a particular orientation of a galaxy, and $\bm{x}$ would be quantities such as the dark matter mass or the tidal field. In this section we briefly introduce two conditional neural density estimators used in this work to model such $p(\bm{y} | \bm{x})$: mixture density networks (MDNs, \citealt{MDN}) and generative adversarial networks (GANs, \citealt{Goodfellow-GAN}). As we will see in the following sections, MDNs will be used to model low-dimensional densities, whereas GANs will be used to model complex joint densities of all galaxies in a halo. We will detail in Sec. \ref{3d_model} how these two models are combined in practice.

\subsubsection{Low-dimensional conditional density fitting with Mixture Density Networks}\label{mlp_mdn}

Mixture density networks are a class of feed-forward neural networks where the outputs are  conditional distributions (the posterior probability distributions) instead of point estimates.  Generally, MDNs are written as the weighted sum of $n_c$ basis PDFs, where we chose  truncated Normal distributions 
as the basis functions: 
\begin{align}
p(\bm{y} | \bm{x} ;\btheta) = \sum_{k=1}^{n_c} r_k(\bm{x} ; \btheta)\, \mathcal{N}\left[\bm{y}\,|\, \boldsymbol{\mu}_k(\bm{x}  ; \btheta),  \boldsymbol{\Sigma}_k(\bm{x}  ; \btheta) , a, b \right]
\end{align}
 where for a given likelihood $p(\bm{y} | \bm{x} ;\btheta)$, the $\btheta$ are all free parameters that are learned during the training, and the
weights $\{r_k(\bm{x}  ; \btheta)\}$, means $\{\boldsymbol{\mu}_k(\bm{x}  ; \btheta)\}$, and covariance matrices $\{\boldsymbol{\Sigma}_k(\bm{x}  ; \btheta)\}$ are dependent on $\bm{x}$ through some neural network. 
Since we are interested in predicting vector quantities that can be represented as unit vectors, 
we used a truncated Normal distribution with $a=-1, b=1$.
To train the MDN, the output at each training step is evaluated for the training dataset and we take the loss function to be the negative log-likelihood:
\begin{align}
L = -\mathrm{log} \left[\prod_{i}  p(\bm{y}_i | \bm{x}_i;\btheta) \right]
\end{align}

\subsubsection{High dimensional conditional density fitting by Adversarial Training}

We now turn to a different strategy for modeling conditional distributions $p(\bm{y} | \bm{x})$ which will become necessary to model the \textit{joint distribution} of multiple properties of all galaxies in a given halo -- in other words, $\bm{y}$ is high dimensional and a MDN will no longer be applicable. To address this problem, we propose to learn this conditional density by adversarial training. 

Given a generating function $g_\theta(z , \bm{x})$ with $z \sim \mathcal{N}(0, \bm{I})$, we aim to adjust the implicit distribution generated by $g_\theta$ to match our target distribution $p(\bm{y} | \bm{x})$. This can be done by minimizing the Wasserstein 1-distance $\mathcal{W}$ between these two distributions to find an optimal set of weights $\theta_{\star}$.
 Unfortunately, we don't have access to $\mathcal{W}$ in closed form, so we must learn it as well. We introduce a second model, such that the expression inside the parnethisis in Eq. \ref{w_dist}
approximates the Wasserstein distance, and which can be trained alongside the generator by solving the following minimax optimization problem:
\begin{equation}\label{w_dist}
	\argmin\limits_{\theta} \left( \sup_\phi \mathbb{E}_{(x, y)} \left[  d_\phi(\bm{x}, \bm{y}) -  \mathbb{E}_{z} \left[  d_\phi(g_\theta(\bm{z}, \bm{x}), \bm{y}) \right]  \right] \right)
\end{equation} 

This is the optimization problem that the WGAN \citep{wgan} aims to solve, modified to account for the conditional variable $\bm{x}$. 
To ensure that $d_\phi$ indeed parameterizes a Wasserstein distance, one must ensure that its Lipschitz constant remains bounded. While several approaches have been proposed, from clipping the weights of the model \citep{wgan}, to applying a spectral norm regularization \citep{sngan}, in this work we adopt the gradient constraint \citep{wgangp} for all applications, as we find it provides good results in practice.

\section{The Simulation and Analysis Methods}\label{methods-section}
 Here we will explain the methods we have used to measure galaxy and DM halos shapes, intrinsic alignment, two point statistics and the methodology we used to decompose/classify galaxies dynamically.

\subsection{Simulated data}\label{simulation}

In this work  we are using the TNG100-1 run from the IllustrisTNG simulation suite \citep[for more information, please refer to][]{ tng-bimodal,pillepich2018illustristng, Springel2017illustristng, Naiman2018illustristng, Marinacci2017illustristng,tng-publicdata}.
The  TNG100-1  is a   hydrodynamical simulation with a box side length of 75 Mpc/h. The simulation uses the moving-mesh code \textsc{Arepo} \citep{arepo} and  contains $2 \times 1820^3$ resolution elements with a gravitational softening length of 0.7 kpc/h 
for dark matter and star particles. 
Within the simulation the dark matter particle mass is $7.46\times 10^6  M_\odot$ and  the star particle masses are variable. Galaxy formation and evolution were modeled using radiative gas cooling and heating; star formation in the ISM; stellar evolution with metal enrichment from supernovae; stellar, AGN and blackhole feedback; formation and accretion of supermassive blackholes \citep{tng-methods}.
The  halos within the simulation were cataloged using friends-of-friends (FoF) 
methods \citep{fof}, and the subhalos  
were cataloged using the SUBFIND algorithm \citep{subfind}. 
The simulation suite includes 100 snapshots at different redshifts, and we use the latest snapshot at $z=0$ for our analysis.  

We employ a minimum stellar mass threshold of $ \log_{10}(M_*/M_\odot) =9 $ for all galaxies, using their stellar mass from  the SUBFIND catalog.   Using the methods described in \cite{gal_decomp}, we quantify the disc fractions of each galaxy based on dynamics (instead of  fitting S\'{e}rsic profiles) and split the sample into two morphological bins: bulge-dominated (those with disc fraction lesser than 0.5) and disc-dominated (those with disc fraction greater than or equal to 0.5). 

\subsection{Tidal field}
In order to predict galaxy and DM halo shapes and orientations, we use the tidal field defined as  the Hessian of the gravitational potential $\phi$:
\begin{equation}
    T_{ij}(\mathbf{r}) = \frac{\partial^2 \phi(\mathbf{r}) } {\partial r_i\partial r_j},
    \label{eq:tidal_field}
\end{equation}

To calculate the gravitational potential $\phi$, we start by computing the matter over-density field by constructing a particle mesh. First, the simulation box is divided into smaller 3D cubic cells. Within each cell $c$ centered at position  $\mathbf{r_c}$, we can count the total mass of the particles in that cell  and divided it by the average across all cells, 
and write the overdensity field as:
\begin{equation}
    \delta (\mathbf{r_c}) = \frac{\rho_c}{ \langle \rho_c \rangle} -1.
\end{equation}

The gravitational potential is related to the over-density field via the Poisson equation:
\begin{equation}
    \nabla^2\phi (\mathbf{r}) = 4\pi G \bar{\rho} \delta(\mathbf{r}),
    \label{eq:gravitational_potential}
\end{equation}
where $G$ is Newton's gravitational constant.
The solution of the Poisson equation in Fourier space is:
\begin{equation}
     \hat \phi(\mathbf{k}) = -4\pi G \bar{\rho} \frac{\hat\delta(\mathbf{k})}{k^2}  ,
\end{equation}

Plugging this result back into the Fourier transform of Eq.~\eqref{eq:tidal_field}, we obtain the tidal tensor 
\begin{equation}
     \hat T_{ij}(\mathbf{k}) = 4\pi G \bar{\rho} \frac{k_ik_j}{k^2}  \hat\delta(\mathbf{k}).
    \label{eq:tidal_field_k}
\end{equation}
In order to smooth the small scale coarseness of the tidal field (caused by the discrete resolution elements of the simulation), we introduce a Gaussian filter with smoothing scale $\gamma$:
\begin{equation}
     \hat{T}_{ij }(\mathbf{k}) = 4\pi G \bar{\rho} \frac{k_ik_j}{k^2} \hat\delta(\mathbf{k}) \;\; e^{-k^2 \gamma^2 /2}.
     \label{eq:tidal_fourier_gaussian}
\end{equation}

Finally, the Fourier-space tidal field from Eq.~\eqref{eq:tidal_fourier_gaussian} can be converted into real space using the inverse Fourier transform.
The tidal field was evaluated at the position of each galaxy, using a cloud-in-cell window kernel to interpolate between the centers of the grid points,
with various values of $\gamma$: 0.1 Mpc/h, 0.25 Mpc/h, 0.5 Mpc/h, 1 Mpc/h and 2 Mpc/h on a mesh of size $1024^3$, with cell sizes given as $L_\text{box}/1024=0.073$ Mpc/h. 

 \subsection{Shapes of Halos and Galaxies}\label{shapes_methods}
 
To measure the shapes of galaxies and DM halos we utilize the mass quadrupole moments (often incorrectly referred to as the inertia tensor%
).  We use the simple quadrupole moments ${ I}_{ij}$,   defined as
\begin{equation}
\label{eq:simple_IT}
{ I}_{ij} = \frac{\sum_n m_n {r_{ni} r_{nj}} }{\sum_n m_n}.
\end{equation}
Here the summation index $n$ runs over all  particles of a given type  
in a given galaxy, where  $m_n$ is the mass of the $n^{\rm th}$ particle and $r$
is the distance between the galaxy or subhalo centre of mass and the $n^{\rm th}$ particle, with $i$ and $j$ indexing the three spatial directions. We chose to use the simple mass quadrupole moment, since   the reduced and the reduced iterative moments generally give very low alignment signals \citep{IA-bulge-disc}.

 The three unit eigenvectors of the mass quadrupole moment, defined   as $\vb s_\mu=\{s_{x,\mu},s_{y,\mu},s_{z,\mu}\}^\tau$ and $\mu \in \{a,b,c\}$, 
  The half-lengths of the principal axes of the ellipsoid are given by $a\propto\sqrt{\omega_a}$, $b\propto\sqrt{\omega_b}$, and $c\propto\sqrt{\omega_c}$, such that $a \geq b \geq c$ and $\omega_a, \omega_b ,\omega_c$ are the   eigenvalues of the mass quadrupole moment.

To compute the projected alignment signals, we need to use the 3D mass quadrupole moments to define 2D projected shapes. Following \cite{joachimi-2013}, we can obtain the projected 2D ellipse by solving $\bm{r}^\tau {\mathbf W}^{-1} \bm{r} = 1$, where 
\begin{equation}
\label{eq:ellipseprojection1}
{\mathbf W}^{-1} = \sum_{\mu=1}^3 \frac{ \bm{s}_{\perp,\mu} \vb \bm{s}_{\perp,\mu}^\tau}{\omega_\mu^2} - \frac{\bm{k} \bm{k}^\tau}{\alpha^2}\;,
\end{equation}
and
\begin{equation}
\label{eq:ellipseprojection2}
\vb k = \sum_{\mu=1}^3 \frac{s_{\parallel,\mu} \bm{s}_{\perp,\mu}}{\omega_\mu^2}\;~~~\mbox{and}~~
\alpha^2 = \sum_{\mu=1}^3 \left( \frac{s_{\parallel,\mu}}{\omega_\mu} \right)^2\;.
\end{equation}
Here, $\vb s_{\perp,\mu} = \{s_{x,\mu},s_{y,\mu}\}^\tau$ are the eigenvectors projected along the projection axis (for which we arbitrarily choose the z-axis of the 3D simulation box). Here, we note that $W$ and $k$ in these equations {\em are not the same quantities} as the previously defined $W$'s and $k$'s in Sections~\ref{DL_intro} and~\ref{methods-section}. 

Then, the  two components of the galaxy ellipticity  can be expressed in terms of the symmetric tensor ${\mathbf W}$ 
\begin{equation}
(e_1,e_2) = \frac{(W_{xx}-W_{yy}, 2W_{xy})}{W_{xx} + W_{yy} + 2 \sqrt{\mathrm{det}\mathbf{W}}}.
\end{equation}
 For the special case that the $\bm{s}_c$ (smallest) axis lies perfectly along the projection axis, the absolute value of the ellipticity is $|e| = (a-b)/(a+b)$. In terms of the projected simulation box, the $x,y$ directions correspond to the positive and negative direction of $e_1$ (since we projected along the $z$ direction).

\subsection{Two-point estimators}
 In this section we will describe the two-point correlation functions that will be used to quantify IA \footnote{All of the two-point statistic were measured using the  \href{https://halotools.readthedocs.io/en/latest/}{HALOTOOLS} package v0.7 \citep{halotools} and the supporting \href{https://github.com/duncandc/halotools_ia}{halotools\_ia} package .}. The ellipticity-direction (ED) correlation captures the position and the orientation correlation angles in 3D which are useful in comparing alignments in  simulations.
 On the other hand, the projected density-shape correlation function ($w_{g+}$) captures the correlation between overdensity and projected intrinsic ellipticity, as is commonly used in observational studies.

\subsubsection{Orientation Correlation Functions in 3D}

The ellipticity-direction (ED) correlation function is defined as \citep{ed-ee}: 
\begin{equation}
\omega(r) = \langle |\hat{e}({\bf x}) \cdot \hat{r}({\bf x})|^2 \rangle -\frac{1}{3}=\omega^{1h}(r) + \omega^{2h}(r)
\end{equation}
for a subhalo/galaxy at position $\bf x$ with major axis direction $\hat{e}$ and the unit vector $\hat{r}$ denoting the direction of a density tracer at a distance $r$. 
As shown, in simulations this correlation function can be decomposed into  1-halo and 2-halo terms, where the 1-halo term captures the correlation among subhalos within the same halo, and the 2-halo term captures contributions from pairs of subhalos belonging to different halos. 
 The estimator we use to compute it in the simulations is as follows:
 \begin{equation} 
   \omega (r) =  \sum_{i \neq j} |\hat{e}_i \cdot \hat{r}_{ij}|^2  - \frac{1}{3}
   \end{equation}

\subsubsection{Density-Shape Correlation Functions in 2D}
The cross correlation function of galaxy positions (as tracers of the galaxy overdensities) and intrinsic ellipticities is defined as:
\begin{equation}
    \xi_{g+} (\mathbf{r} ) = \langle \delta_g(\mathbf{x}) \delta_+(\mathbf{x}+\mathbf{r}) \rangle
\end{equation}
where $ \delta_g(r)$ and $\delta_+(r)$ represent the galaxy over-density field and the intrinsic shape field, respectively. It  can be estimated using the method described in \citet{mandelbaum-2011} 
as a function of $r_\mathrm{p}$ and $\Pi$:
\begin{equation}
\xi_{g+} (r_\mathrm{p}, \Pi) = \frac{S_+D - S_+R}{RR}.
\end{equation}
 Here,  $RR$ are counts of  random-random  pairs  binned based on their perpendicular and line-of-sight separation; 
\begin{equation}
S_+D \equiv \frac{1}{2} \sum_{\alpha\neq \beta} e_{+}(\beta|\alpha),
\end{equation}
 represent the shape correlations,  
 where   $e_{+}(\beta|\alpha)$ is the $+$ component of the ellipticity of galaxy $\beta$ (from the shape sample) measured relative to the direction of galaxy $\alpha$ (from the density tracer sample). $S_+R$ is defined in an equivalent way, but instead of using galaxy positions we use randomly distributed positions.

 The projected two-point correlation functions can be obtained by integrating over the third dimension, with the integral approximated as sums over   the line-of-sight separation ($\Pi$) bins:
 \begin{equation}\label{eq:xi_to_w_sum}
w_{g+} (r_\mathrm{p}) = \sum_{-\Pi_\mathrm{max}}^{\Pi_\mathrm{max}} \Delta\Pi \,\xi_{g+} (r_\mathrm{p}, \Pi),
\end{equation}
where we chose a  $\Pi_\mathrm{max}$ value of 20 Mpc/h, following \cite{IA-bulge-disc}.

\section{Neural Intrinsic Alignment Model}\label{neural_model}

Having introduced the fundamental machine learning concepts necessary to build our model, as well as the necessary background on simulation data and intrinsic alignments, we now introduce our proposed model.

\subsection{Graph construction}\label{graph_const} 
 
To construct the graph for the cosmic web (i.e., for the subhalos and the galaxies), we first  grouped all of the subhalos and galaxies based on their parent halo. In other words, we grouped subhalos and galaxies based on their group membership ID from the halo finder. There exist a few choices for modeling proximity graph relations between the members in a group, such as the Gabriel graph (which is a subset of the Delaunay triangulation) and the radius nearest neighbor graph  (r-NNG).  These different graph types  differ by their
connectivity, i.e., they have different adjacency matrices \citep{proximity}. In this study we employ the r-NNG to model the connectivity of our graphs.

 Given a  galaxy catalog,  an undirected graph based on
the 3D positions is built by placing each galaxy on a \textit{graph node}.   Then, each node will have a list of features such as halo mass, subhalo mass, central vs.\ satellite identification (binary column) and tidal fields smoothed on several scales.   
Then for a given group (i.e., within a halo) the graphs are connected. To build the graph connection,  the  nearest neighbors within a specified radius for a given node are connected via the \textit{undirected edges} with \textit{signals} on the graphs representing the alignments.

\subsection{Model Architecture}\label{general_arch}
 In Fig.~\ref{diag} we outline the architecture of our models. The general idea is that we have list of features (orange box) that are relevant for capturing the dependence of intrinsic alignments within a halo (dashed red box), and the tidal fields that are relevant for capturing the dependence of IA for galaxies on matter beyond their halo
 (dashed purple). Then, these inputs are fed into the GAN-Generator (crimson box), which tries to learn the desired output labels (yellow box). At the end the input and the output from the GAN-Generator are fed into the GAN-Critic (blue box) to determine the performance of the GAN-Generator. 
 In our model, the Generator has 5 layers each with \{128,128,16,2,2\} neurons, while the Critic has 4 layers each with \{128,128,64,32\} neurons followed by a mean-pooling layer and a single output neuron. For scalars, this model works as is. 
 
 However, for predicting vector quantities, we made slight modifications to the network architecture.
 For the 2D model, the input tidal fields are all in 3D, but the GAN-Generator outputs a 2D vector, while the inputs of the GAN-Critic get projected onto 2D. 
 For 3D vector quantities (orientations), we had to slightly modify our architecture as we will explain below.

\subsubsection{3D orientation modeling}\label{3d_model}
\begin{figure*} 
\includegraphics [width=6.in]{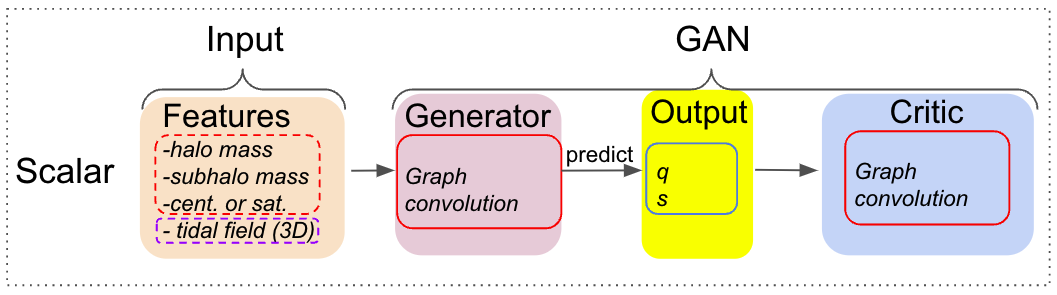}
\includegraphics [width=6.in]{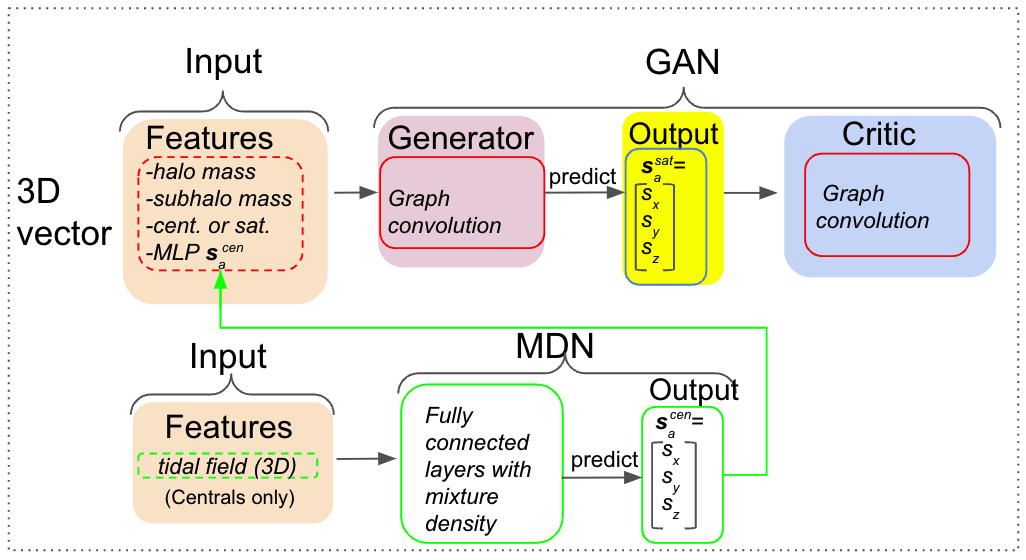}
\includegraphics [width=6.in]{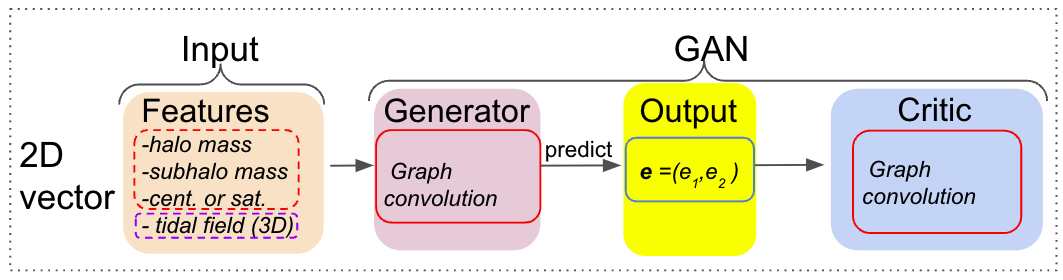}
 \caption{Architecture of the Graph convolution GAN models used. 
 }\label{diag}
 \end{figure*}

Since satellite alignments are challenging to model, we rely on the \textit{graph} structure to capture this alignment, whereas for the centrals   we use a simple  MDN.  The structure of the network in this case is shown in the ``3D vector'' box within Fig.~\ref{diag}. We decided to take a `divide-and-conquer' approach where we broke down the problem as follows:
\begin{equation}
    p(\bm{y}| \bm{x}) = p(\bm{y}_\text{cen}| \bm{x}_\text{tid}) \, p(\bm{y}_\text{sat}|\bm{y}_\text{cen}, \bm{x}_\text{sat})
\end{equation}
where $p(\bm{y}| \bm{x})$ represents the conditional probability of the  3D orientations of all galaxies given all input features; $p(\bm{y}_\text{cen}| \bm{x}_\text{tid})$ is the 3D orientation of central galaxies given the tidal fields;
and $p(\bm{y}_\text{sat}|\bm{y}_\text{cen}, \bm{x}_\text{sat})$ is the 3D orientation of satellite galaxies given the features of satellite galaxies and orientations of central galaxies.  $p(\bm{y}_\text{cen}| \bm{x}_\text{tid})$ is modeled using the tidal fields with MDN as shown in the transparent green box in the second panel of Fig.~\ref{diag}.  Next, the outputs from the MDN are fed into the GAN, together with the other features, as listed in the red dashed box (we feed the output from the MDN to the GAN in order to capture correlations between the centrals and satellites). 
Since the satellite alignments appeared to be modeled effectively 
without using the tidal fields as an input feature, we did not include the tidal fields in the input for the GAN. 
Note that we initially tried training the GAN  to capture both central and satellite alignments, but it was under-predicting the alignments for centrals, which motivated the approach described above.

\subsection{Training}
 We train the model using the Adam optimizer \citep{adam} 
 with a learning rate of $10^{-3}$ and exponential decay rates of $\beta_1 = 0$ and  $\beta_2 = 0.95$. During the adversarial training we train the Generator for 5 steps and the Critic for 1 step with  a batch size of 64 (one batch is set of graphs) and a leaky ReLU activation function. Due to the scarcity of high mass halos, we balance probabilities of graphs based on group mass when processing batches by down-sampling halos of low mass and up-sampling halos of high mass with replacement. As a result, halos with high mass are reused in multiple different batches in the training.

In order to avoid overfitting, during each training epoch we augment the data by applying random rotations to the batches of the graphs.  As a test for overfitting, we also made roughly a 50/50 train-and-test sample split, while still maintaining group membership. The results of this test are in Appendix \ref{appendix:overfit}; we see no significant signs of overfitting. Given these findings and the limited simulation data available, in the following presentation of results, we used all of the sample to generate the output.

 As is common with GANs, our GAN models do not converge; we had to arbitrarily stop the training once it reached a reasonable result. For example, for the 2D model the results start to look reasonable around training step 40000, which takes about 3--4 days on a single NVIDIA-A100 GPU. 
 Our code is available at \url{https://github.com/yesukhei-git/GraphGAN}.

\section{Results  }\label{results}
In this section, we will first describe the results when using the GAN to predict scalar quantities, then 3D orientation correlation functions, and finally the 2D (projected) intrinsic alignment correlations. 

Throughout the section we will refer to the sample generated from the Graph-Convolutional Network-based Generative Adversarial Networks as the \textit{GAN} sample, and the samples from the TNG100 simulation as the \textit{TNG} sample. In all of the below, we always trained and generated on the full sample, and tests on the subsamples were used to demonstrate the ability of the models to distinguish the variation in IA across the subpopulations. Since past work \citep[e.g.,][]{IA-bulge-disc} has shown that alignments
depend primarily on mass, satellite/central status and morphology,
we explicitly check that the GAN can capture these dependencies.

\subsection{Predictions of scalar quantities: shapes}
In this subsection, we present the   shapes of DM subhalos and galaxies.  Shapes are important scalar quantities that is used in intrinsic alignment studies. The DM subhalo and galaxy shapes were generated separately. The model for predicting DM subhalo shapes serves a dual purpose: it can be used as a sanity check and used to predict shapes of subhalos that are not well resolved.  

\begin{figure*} 
\includegraphics [width=7.in]{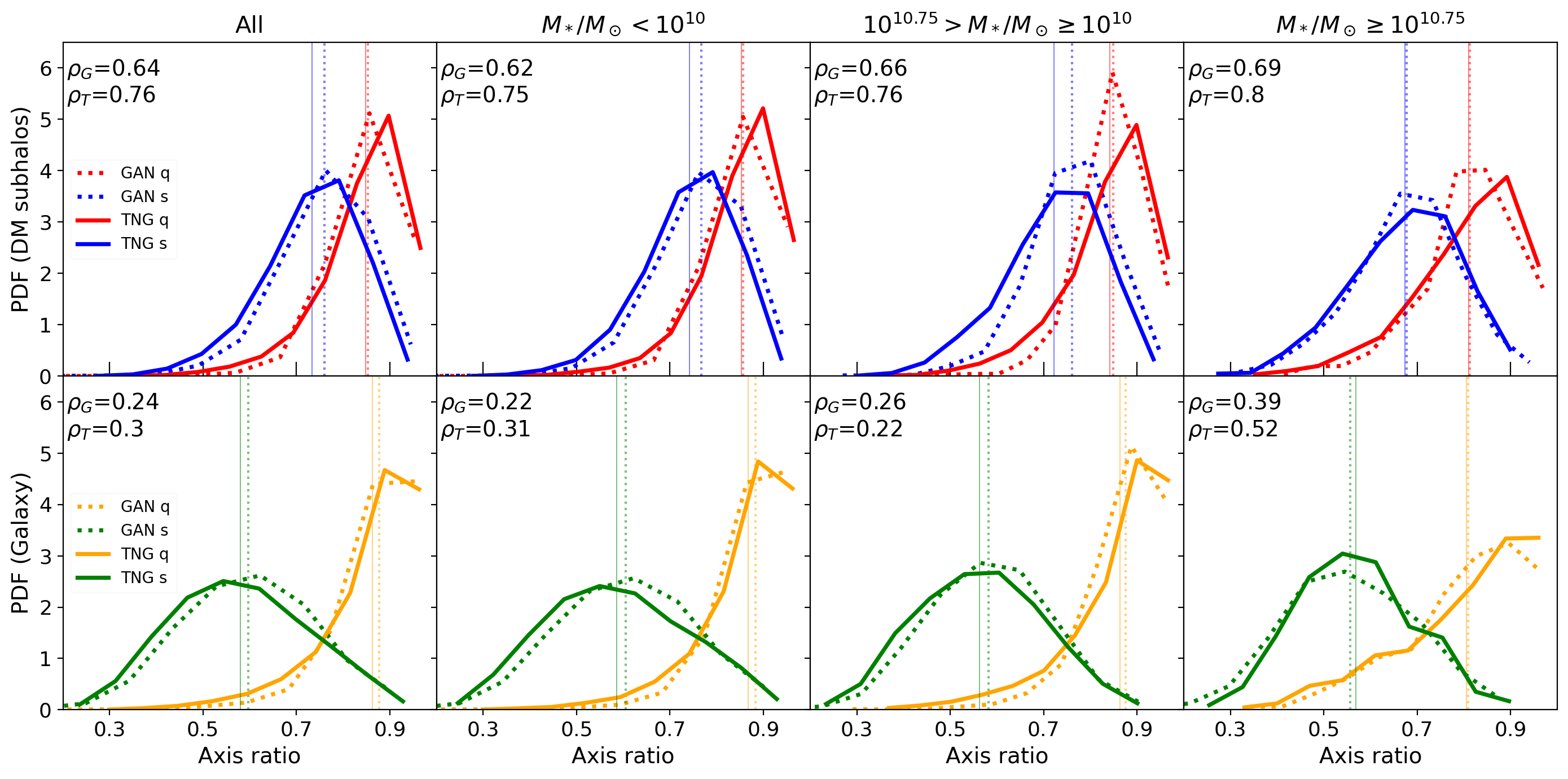}
\includegraphics [width=7.in]{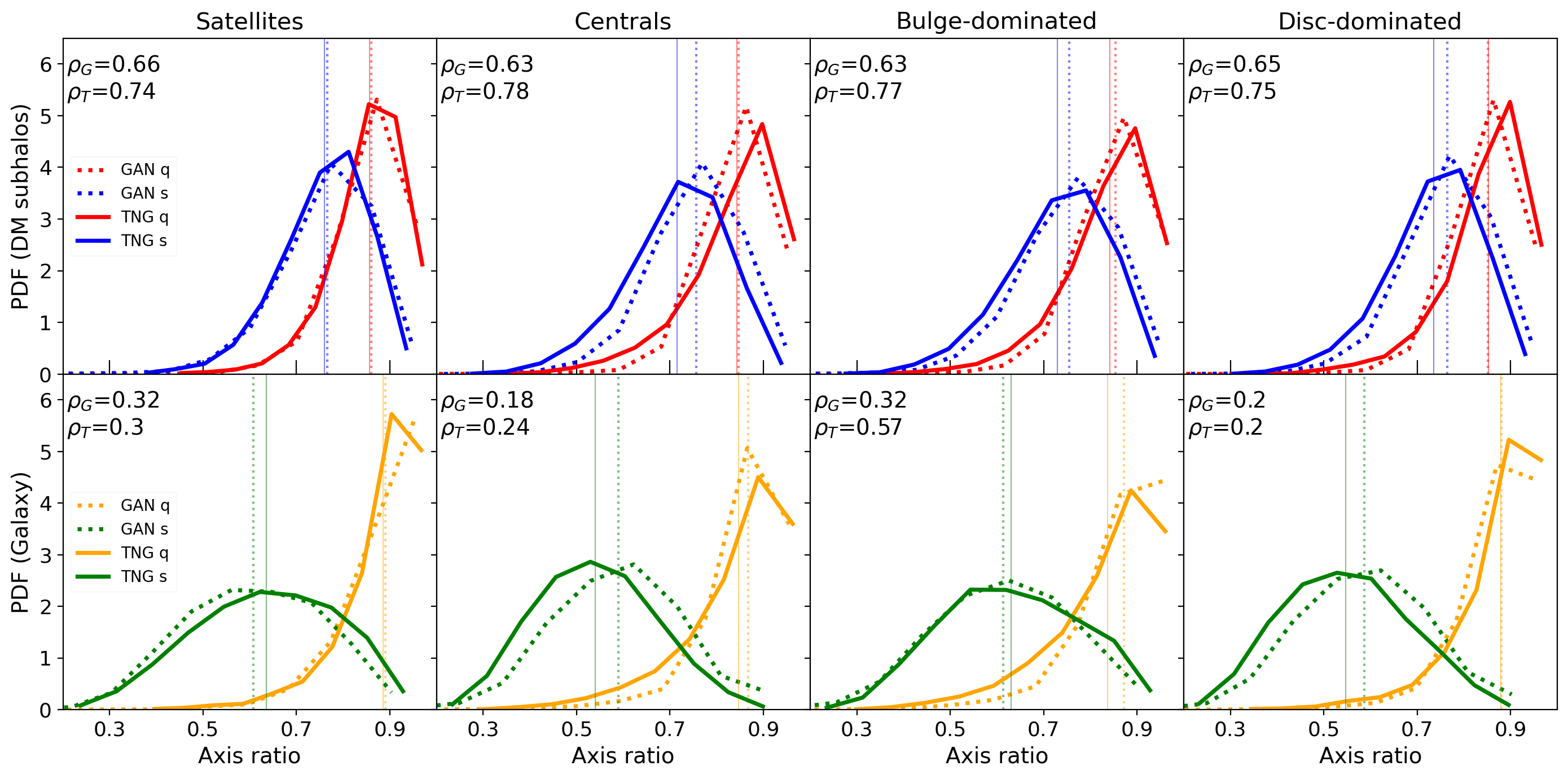}
 \caption{Distribution of shapes of DM subhalos (top row) and galaxies (bottom row), quantified using the intermediate-to-major axis ratio $q$ and the minor-to-major axis ratio $s$ for the galaxy or subhalo modeled as a 3D ellipsoid.  We show results for the full sample, and for subsamples determined based on stellar mass, satellite/central identification, and morphological classification. Solid lines indicate the sample from the TNG simulation and the dotted lines indicate the sample generated by the GAN. The parameters $\rho_G = \rho(\text{GAN} \, q, \text{GAN} \, s)$ and $\rho_T = \rho(\text{TNG} \, q, \text{TNG} \, s)$ in each panel indicate the Pearson-r correlation between the two shape parameters.
 The faint vertical lines indicate the mean of the distribution with the corresponding style and color.
 Qualitatively, the GAN reproduces the shape of the distributions for the whole sample and the various subsamples. Quantitatively, the GAN captures the mean and the spread of the distribution well. However, the GAN typically underestimates the correlation between $q$ and $s$ by about 5--20 per cent for a given subsample. Note: for DM subhalos we used the corresponding galaxy masses for the binning.}\label{shapes}
 \end{figure*}

In Fig.~\ref{shapes} we compare the histograms of galaxy and DM subhalo axis ratios, defined as $q=b/a$ (intermediate-to-major axis ratio) and $s=c/a$ (minor-to-major axis ratio).  
Overall, the GAN captures and reproduces the distributions of the two axis ratios to a good degree, with the means of distributions agreeing  within a few percent. 
We also measured the correlation between the two shape parameters using the \textit{Pearson-r} coefficients, which are displayed in each panel as $\rho$.  These values tell us to what degree the GAN captures the distribution of 3D shapes from TNG (since the 3D shape distribution depends on correlations between $q$ and $s$ within the population).  For the DM subhalo shapes, the GAN underestimates the correlation between the two shape parameters by about 10-15 per cent for all the subsamples. In contrast, for the galaxy shapes, the GAN underestimates the correlation between the two shape parameters by about 5-20 per cent for all  subsamples, except for the satellites where it is within 2 per cent of the target correlation. All in all, the distributions of shapes generated by the GAN are in good quantitative agreement with the target distributions, though slightly underestimating the correlation between the two shape parameters. 

In contrast,  the correlation defined as $\rho(\text{GAN} \, q, \text{TNG} \, q)$  -- and likewise for $s$ -- is less informative, because we care about the population statistics, and not accurate predictions for individual subhalos/galaxies.  Indeed, a high value of this correlation coefficient would be a bad sign, as it might imply that the model has `memorized' the specific shapes of galaxies/subhalos in TNG, rather than learning about the distribution of shapes for the ensemble.    We find that the measured $\rho(\text{GAN} \, q, \text{TNG} \, q)$ and $\rho(\text{GAN} \, s, \text{TNG} \, s)$   were below 15 per cent for all subsamples. These numbers are not shown on the figure.

\subsection{Predictions of vector quantities}

\begin{figure*}\label{ellip}
\begin{subfigure}{.4\textwidth}
\centering
\includegraphics[width=8.5cm ]{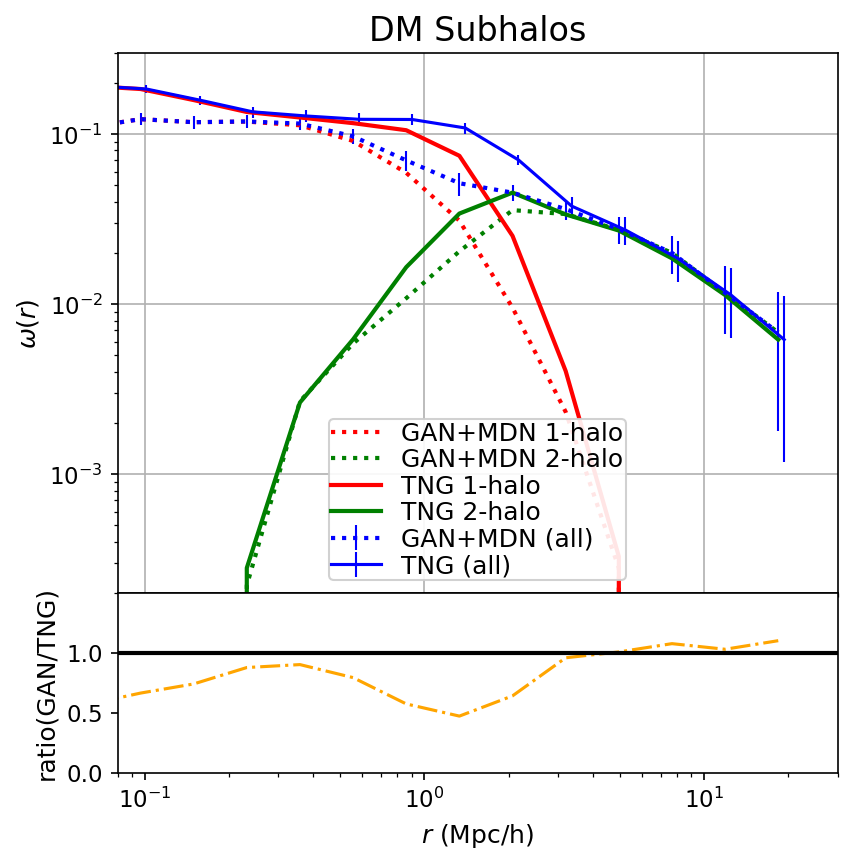}
 
\end{subfigure}\hfill
\begin{subfigure}{.49\textwidth}
\centering
\includegraphics[width=8.5cm ]{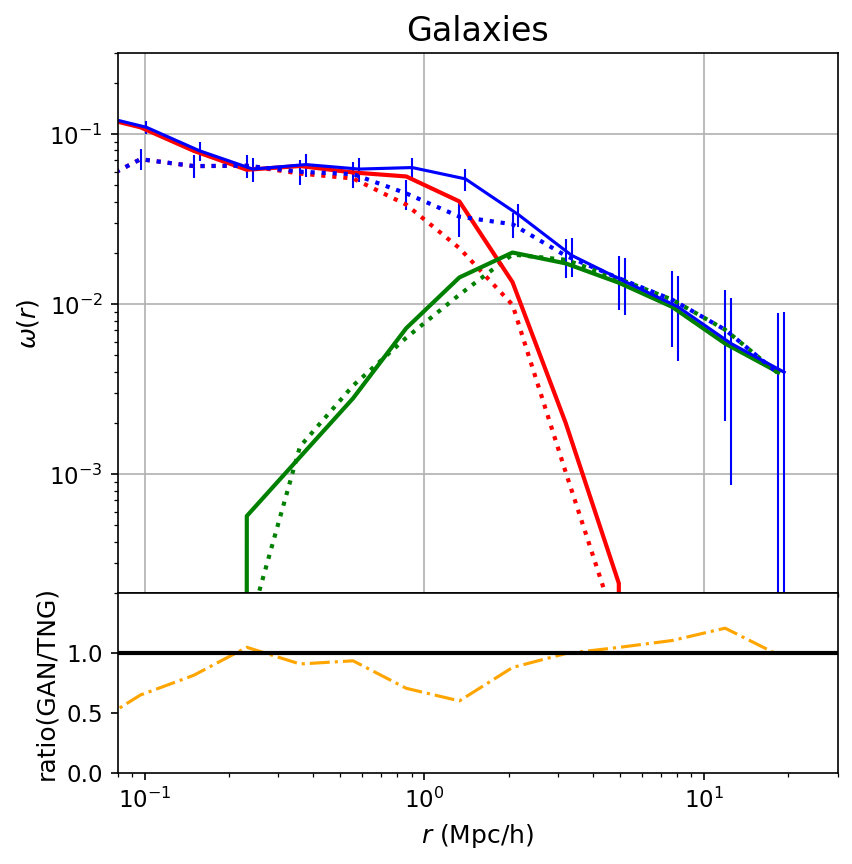}
 
\end{subfigure}\hfill
 
\caption{\label{ED_all}  
 ED correlation function, $\omega(r)$, of the 3D major axis with galaxy positions: the solid lines show the measured values from the TNG simulation, while the dashed lines show the generated values from the GAN+MDN. The top panels show  $\omega(r)$ decomposed into 1-halo (red line) and 2-halo (green line) terms along with the total (blue line), and the bottom panel shows the ratio of the total $\omega(r)$ from the GAN+MDN to that measured in TNG. The panel on the left is for DM subhalos, whereas the right panel is for galaxies. The generated values agree well with measured values on most scales, except on very small scales ($\lesssim$0.1 Mpc/h) and the 1- to 2-halo transition region ($\sim$1 to 2 Mpc/h). The GAN+MDN curve was shifted by 5 per cent to the left for visual clarity.
}
\end{figure*}

\begin{figure*}
\centering
 {\includegraphics[width=0.95\textwidth]{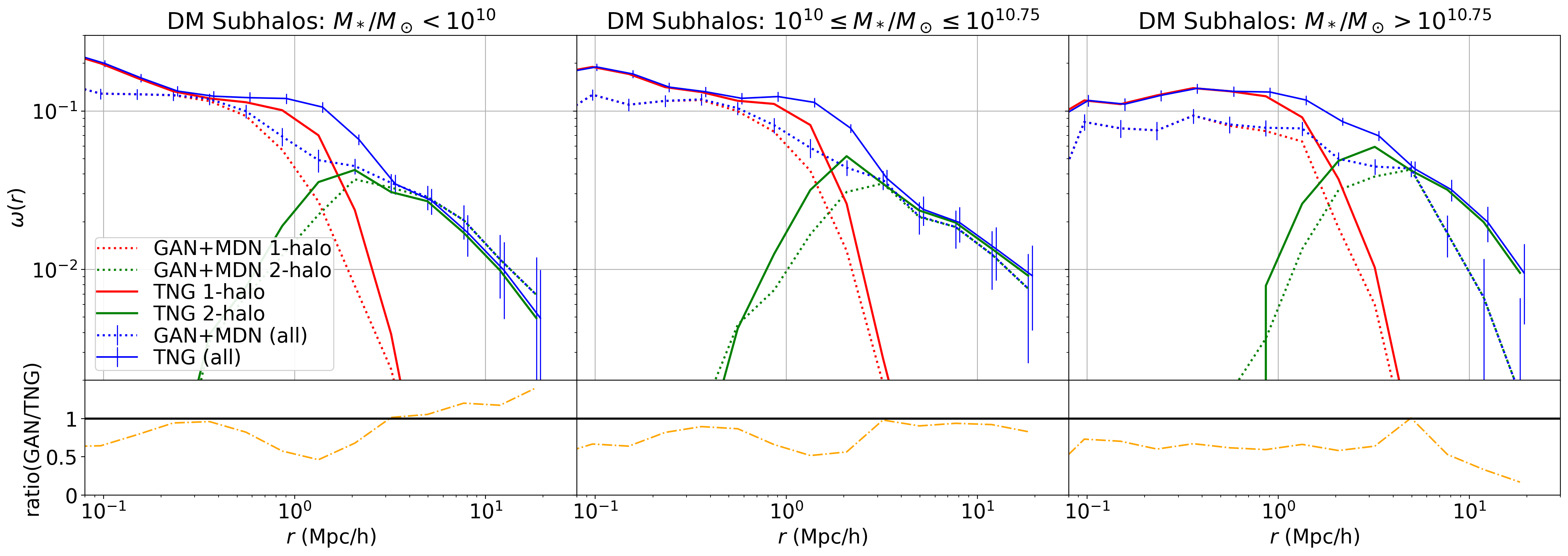}}%
\\ 
{\includegraphics[width=0.95\textwidth]{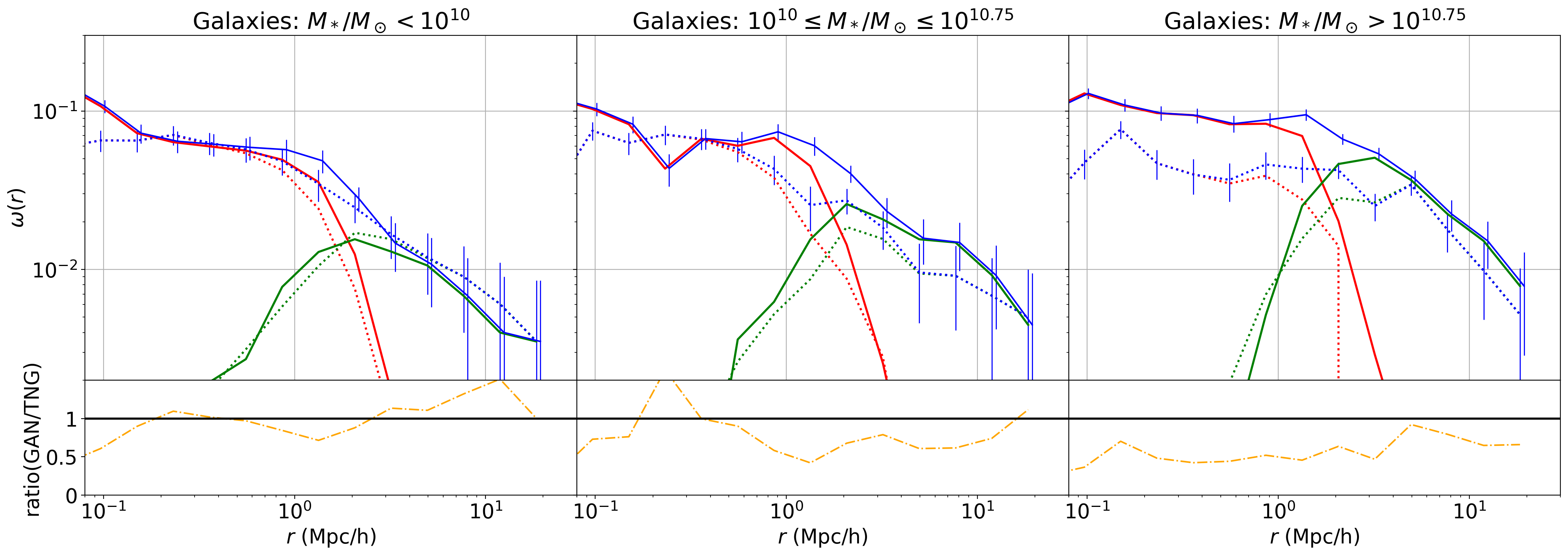}}%
\caption{\label{ED_mass}  
 Same as Fig.~\ref{ED_all}, but after dividing the sample into mass bins as indicated on top of each figure. 
 For both DM subhalos and galaxies, for the two lower mass bins, the agreement between the generated model and TNG is good on most scales and follows the same trend as the whole population (see Fig.~\ref{ED_all}). However, for the highest mass subsample, for both DM subhalos and galaxies, the generated values from the GAN+MDN underestimate the alignment correlation function by a factor of $\sim 1/2$ on most scales. This may be due to the fact that there are far fewer high mass galaxies than low mass galaxies in our sample, resulting in an insufficient training sample for that mass range. Note: for DM subhalos we used the corresponding galaxy masses for the binning.
}
\end{figure*}

\subsubsection{Predictions of ED correlation functions  }
Here, we present the results of the  correlation functions of the galaxy major axis directions with galaxy position in 3D. In Fig.~\ref{ED_all}, we plot the ED correlation function for DM subhalos and galaxies for the whole population generated from the GAN.  The error bars were obtained using jackknife resampling. Here we remind the reader that the alignments of satellites were learned using the Graph-structure and the alignments of centrals were learned using a simple  MDN, as described in \ref{mlp_mdn}. Also, the output from the MDN was given as an input to the Graph-structure so as to capture any correlation between the central and satellite terms. 
The satellite alignments are generally more challenging to model, since they are strongly affected by physics on nonlinear distance scales with many competing physical processes that shape the alignment \citep{ia-review2}, 
and the Graph-structure learns and captures the satellite alignment correlations using only mass for both the DM subhalo and the galaxies. The alignment of centrals  is easier to model, so for simplicity and efficiency we learned it using a simple  MDN regressor. 
The generated alignment correlations provide a reasonable match to the simulation on many scales.  The exception is that it  underestimates the correlation (by a factor of 0.5) on $\lesssim$0.1 Mpc/h scales, and in the 1- to 2-halo transition region, $\sim$1--2 Mpc/h. 

Next, in Fig.~\ref{ED_mass}, we present similar quantities as in Fig.~\ref{ED_all} for subsamples based on mass.  For this purpose, we use mass bins of $M_*/M_\odot <10^{10}, 10^{10 } \leq M_*/M_\odot \leq 10^{10.75}, M_*/M_\odot > 10^{10.75}$; for DM subhalos, we used the corresponding galaxy masses for the binning. For the low and intermediate mass bins (the first two columns), the GAN-predicted curve follows a very similar pattern as in Fig.~\ref{ED_all}, with good agreement on most scales. However, for the high mass bin (third column), the GAN underestimates the alignment correlation function by about a factor of two on all scales, for both DM subhalos and galaxies.   This under-performance may be explained by the small number of high mass galaxies in our sample,  which may have caused the neural network to be undertrained for this mass bin. A similar effect has been observed in \cite{ho-halo-mass} where the support density machine was also under-performing for high mass halos.

   \begin{figure} 
   \centering
\includegraphics [width=2.7in]{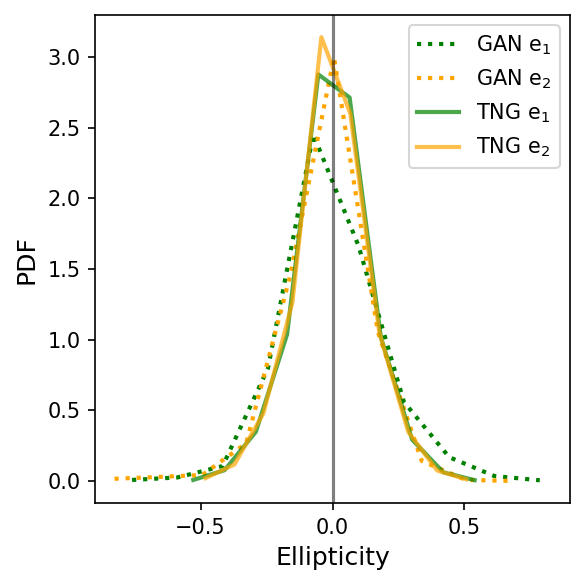}

 \caption{Distribution of  individual components of the 2D complex ellipticities. 
 The measured values from  TNG100-1 are shown with solid lines, whereas the GAN-generated ellipticities are shown with dotted lines, with green denoting $e_1$ and yellow denoting $e_2$. The grey vertical line serves as a reference point  of zero. The GAN produces distributions of ellipticity values that agree quantitatively with the distributions measured directly from TNG.  The distributions have means of -0.02, 0.00, -0.01, 0.00 and standard deviations of 0.17, 0.15, 0.14, 0.14 for GAN $e_1$, GAN $e_2$, TNG $e_1$, TNG $e_2$, respectively.  
 }\label{e1e2}
 \end{figure}

    \begin{figure*} 
\includegraphics [width=6.8in]{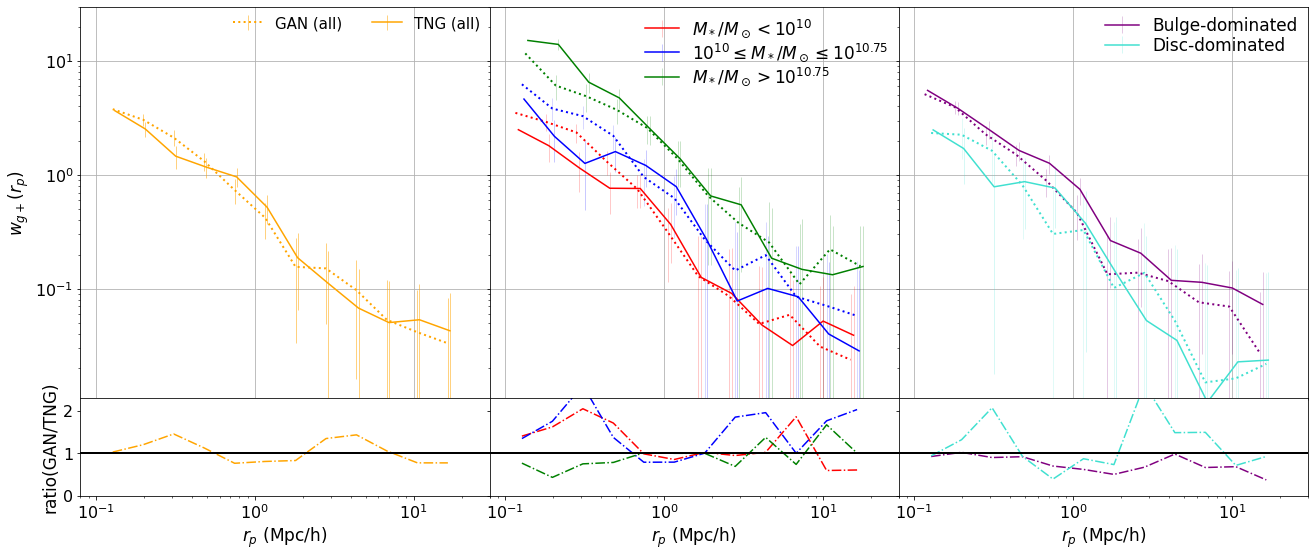}
 \caption{The projected two-point correlation functions $w_{g+}$ of galaxy positions and the projected 2D ellipticities of all galaxies (first column), for subsamples divided by mass (second column) and for two morphological subsamples (third column), as indicated in the legends. The top rows show  $w_{g+}$ measured using data from the TNG simulation in solid lines and the data generated by the GAN in dotted lines, the bottom   panel shows the ratio of the GAN curve to the TNG curve. For the first column, the two curves are in good quantitative agreement an all scales, with ratios closely following 1. For the second column, qualitatively the GAN captures the trends with mass, with higher mass bins showing higher alignment signal. Quantitatively, there is good agreement down to 0.5 Mpc/h scales. At scales below 0.5 Mpc/h the GAN underestimates the signal for the highest mass bin, whilst overestimating the signal for the lower two mass bins. For the third column, the GAN correctly reproduces the different intrinsic alignment for the two morphological types, even though this information was not explicitly given, whilst maintaining a quantitative agreement with the measured values. The errors are dominated by large-scale structure and may be correlated between the curves shown for TNG and the GAN, rather than being independent.
 }\label{wgp_panels}
 \end{figure*}

    \begin{figure*} 
\includegraphics [width=6.7in]{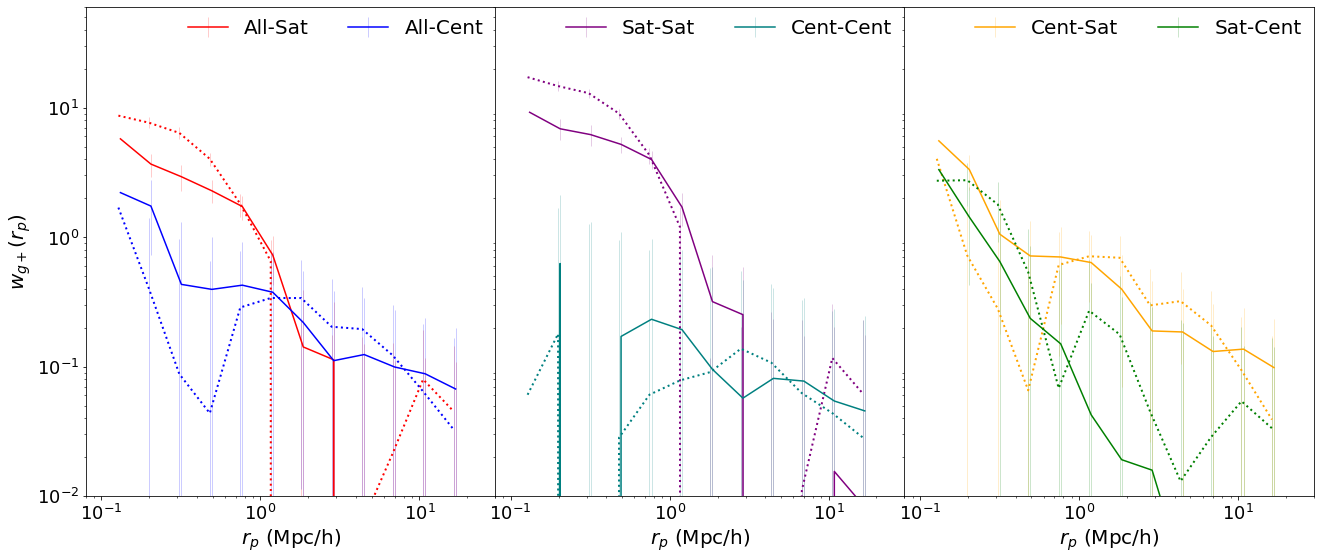}
 \caption{Projected two-point correlation function $w_{g+}$ of galaxy positions and the projected 2D ellipticities, for subsamples defined by their central-satellite distinction.   The plot shows $w_{g+}$ measured using data from the TNG simulation in solid lines and the data generated by the GAN in dotted lines The labels specify which samples were used as the density tracers (first word) and shape samples (second word), e.g., `All-Cent' refers to the correlation between the shapes of central galaxies and the positions of all galaxies. 
 Qualitatively, the GAN captures the distinction between centrals and satellites. We excluded the ratio panels in this figure because some $w_{g+}$ values were zero, and we only wanted to present the qualitative agreement.  
 }\label{wgp_cent_sat}
 
 \end{figure*}
\subsubsection{Predictions of $w_{g+}$ correlation functions    }
Moving to 2D shapes and alignments, in Fig.~\ref{e1e2} we present the distribution of 2D complex ellipticities from the TNG simulation and the GAN. The GAN produces distributions of ellipticities that agree well quantitatively with the ones measured from the TNG simulation, with   means of -0.02, 0.00, -0.01, 0.00 and standard deviations of 0.17, 0.15, 0.14, 0.14 for GAN $e_1$, GAN $e_2$, TNG $e_1$, TNG $e_2$, respectively. 

Next, we examine $w_{g+}$, the density-shape correlation function  computed using the ellipticities, as shown in Fig.~\ref{wgp_panels}. Compared with the 3D correlation function, $\omega(r)$, the projected 2D correlation function, $w_{g+}$, from the GAN agrees quantitatively with the measured one from TNG simulation. 
Here, the errorbars were derived from  an analytic estimate of the covariance matrix, which includes Gaussian terms for  noise and cosmic variance (for more details see \citealt{singh-covarience,samuroff-2020}). We explore the robustness of the models and their stochastic uncertainties in Appendices~\ref{appendix:overfit} and~\ref{appendix:aleatoric}, respectively. These  results show no sign of overfitting, and the stochastic uncertainty of the model is substantially smaller than the analytic errors representing astrophysical sources of statistical uncertainty (shape noise and cosmic variance). 

In the second column of Fig.~\ref{wgp_panels},  we examine the mass dependence 
of the $w_{g+}$ curve from the GAN. Again, here the agreement between the GAN and the measured correlation functions in TNG are quite good for all three mass bins. Still, on scales around 0.1--0.5 Mpc/h, the GAN overestimates $w_{g+}$ for the low and intermediate mass bins, while underestimating $w_{g+}$ for the high mass bin by about a factor of 2. This again may be due to the fact that there are far fewer high mass galaxies to train on for the GAN. 

One thing to note here is that this plot shows the GAN model trained on the total mass of the subhalos, instead of separate columns for DM, gas and stellar mass. However, when we give the GAN DM, gas and stellar mass information, the predicted mass dependence of the intrinsic alignments improves significantly, with disagreement only on the very smallest scales. We nonetheless chose to show the GAN model trained on total mass, since we plan to deploy this model on N-body simulations where that is the only information available. 

Next, in the third column of Fig.~\ref{wgp_panels}, we explore whether the GAN captures the alignment dependence on galaxy morphology. 
Interestingly, the GAN is able to distinguish between morphologically selected samples, even though it was never given this information explicitly. As expected, the bulge-dominated sample shows higher alignment signal in both TNG and GAN, with good quantitative agreement. Similarly as expected, the disc-dominated population shows lower alignment, again with good quantitative agreement. Note that the bulge-dominated sample was controlled for mass effects (i.e. this sample was weighted to match the mass distribution of the disc-dominated sample), as explained in \cite{IA-bulge-disc}. Therefore the implicit mass differences between bulges and discs have already been accounted for in this comparison. 
We hypothesize that the network may be able to distinguish between morphological types based on  the environments through the tidal fields, since galaxy morphology tend to depend on the environment \citep{env-sdss}.
We note that this morphological split is not available for N-body simulations, since it is based on the kinematics of the galaxy (unless galaxy morphologies are included as part of the model).

Finally, we quantify whether the GAN captures the different intrinsic alignments of central versus satellite galaxies. 
In Fig.~\ref{wgp_cent_sat}, we present the $w_{g+}$ measured using different combinations of central and satellite galaxies for the density and shape samples, as indicated by the labels. The figure shows that the GAN-generated $w_{g+}$ follow similar trends as the TNG measured $w_{g+}$ when distinguishing between central and satellite alignments. However, the results may differ at the level of a factor of two on  small scales for the All-Sat, Cent-Cent and Sat-Cent curves, which exhibit higher alignment signals, with the GAN overestimating the alignments. For the All-Cent, Sat-Sat and Cent-Sat curves the GAN also captures the trend, and even though it is in quantitative agreement with TNG, we note that these signals are very low and noise dominated.

\section{Conclusions}\label{conc}
 
 In this article, we have developed a novel deep generative model for intrinsic alignments. Using the TNG100 hydrodynamical simulation from the IllustrisTNG simulation suite, we have trained the model to accurately predict 3D shapes and the 3D orientations of the major axess of both DM subhalos and galaxies; and the projected 2D complex ellipticities of galaxies. For a simulation box of 75 Mpc/h with 20k galaxies it takes about 3--4 days on a modern GPU to train; applying the model on a dataset of equal size is very fast (less than a minute). 
 
 For the scalar quantities -- the shapes of DM subhalos and galaxies -- the GAN model generated values that were in good quantitative agreement with the distributions of their actual measured counterparts from the TNG simulation. However, the correlation between the two shape parameters (intermediate-to-major and minor-to-major axis ratio) was slightly underestimated, by around 5--15 per cent depending on the galaxy or subhalo subsample.
 
 Next, for the 3D vector quantities, the GAN generated the 3D major axis for the whole sample, and the resulting alignment correlation functions agree well for most scales except the very small and the 1- to 2-halo transition scales. When considering mass-selected subsamples, the results for small and medium mass subhalos and galaxies are similar to those of the whole sample. However, for the highest mass bin, the GAN-generated values underestimate the alignment correlation function by about a factor of 2 on most scales.
 
 Finally, for the projected 2D complex ellipticities, the projected density-shape correlation functions ($w_{g+}$) compused using the GAN-generated ellipticities are in excellent quantitative agreement with those from TNG100. Even when considering mass-selected subsamples, the quantitative agreement is good, except for scales below 0.5 Mpc/h where the high mass $w_{g+}$ tends to be underestimated, while $w_{g+}$ for intermediate and lower mass subsamples is overestimated by about a factor of 2. Also, the can qualitatively capture the IA trend for centrals and satellites, and for morphology-selected subsamples.
 
 Overall, the Graph Convolution based Generative Adversarial network learns and generates scalar and vector quantities that have statistical properties (distributions and alignment correlations) that agree well  with those of the simulation. The primary deficiencies were in high mass subsamples, perhaps due to insufficient training data in that regime.

In the future, we would like to deploy this model on a much
higher volume N-body simulation with lower resolution in order
to fully harness its power for upcoming weak lensing surveys. For
this purpose, we will need the features listed in Fig. 1, which are
usually available for N-body simulations.  
Additionally, in order to test the sensitivity of the GraphGAN performance to the subgrid physics implementation in the simulation used for training, we want to apply the trained model to a different high-resolution simulation with a different galaxy formation model, as well as to a low-resolution N-body simulation with a different gravity solver.

 As another area for future work, when we modeled the 3D orientation we only did so for one axis of the ellipsoid. 
An interesting direction for future work is to use SO(3) or E(3) equivariant neural networks for graphs \citep{iso-gcn,egnn} and point sets \citep{thomas2018tensor,villar2021scalars}. 
However,  if one wants to use a light cone (which will need additional training across redshifts), or use 2D alignment statistics within a snapshot, then a new implementation of SO(3) equivariant neural networks is not needed. 
 Given an N-body simulation with the correct features, our model can be used to include IA for very little additional computational cost.  
  \section*{Data Availability}
 
 The data used in this paper is publicly available.
The IllustrisTNG data can be obtained through the website at
\url{https://www.tng-project.org/data/}. The catalog data with morphological decompositions of galaxies is available at \url{https://github.com/McWilliamsCenter/gal_decomp_paper}.
The software developed as part of this work is available at \url{https://github.com/yesukhei-git/GraphGAN}.

\section*{Acknowledgements}

We thank  Ananth Tenneti, Tiziana DiMatteo, Barnabas Poczos and Rupert Croft for useful discussion that informed the direction of this work.  We acknowledge the anonymous referee for feedback that improved the presentation of this work. This work was supported in part by the National Science Foundation, NSF AST-1716131 and by a grant from the Simons Foundation (Simons Investigator in Astrophysics, Award ID 620789). SS is supported by a McWilliams postdoctoral fellowship at Carnegie Mellon University. This work is supported by the NSF AI Institute: Physics of the Future, NSF PHY- 2020295.

\bibliographystyle{mnras}
\bibliography{example}

\begin{thebibliography}{}
\makeatletter
\relax
\def\mn@urlcharsother{\let\do\@makeother \do\$\do\&\do\#\do\^\do\_\do\%\do\~}
\def\mn@doi{\begingroup\mn@urlcharsother \@ifnextchar [ {\mn@doi@}
  {\mn@doi@[]}}
\def\mn@doi@[#1]#2{\def\@tempa{#1}\ifx\@tempa\@empty \href
  {http://dx.doi.org/#2} {doi:#2}\else \href {http://dx.doi.org/#2} {#1}\fi
  \endgroup}
\def\mn@eprint#1#2{\mn@eprint@#1:#2::\@nil}
\def\mn@eprint@arXiv#1{\href {http://arxiv.org/abs/#1} {{\tt arXiv:#1}}}
\def\mn@eprint@dblp#1{\href {http://dblp.uni-trier.de/rec/bibtex/#1.xml}
  {dblp:#1}}
\def\mn@eprint@#1:#2:#3:#4\@nil{\def\@tempa {#1}\def\@tempb {#2}\def\@tempc
  {#3}\ifx \@tempc \@empty \let \@tempc \@tempb \let \@tempb \@tempa \fi \ifx
  \@tempb \@empty \def\@tempb {arXiv}\fi \@ifundefined
  {mn@eprint@\@tempb}{\@tempb:\@tempc}{\expandafter \expandafter \csname
  mn@eprint@\@tempb\endcsname \expandafter{\@tempc}}}

\bibitem[\protect\citeauthoryear{{Arjovsky}, {Chintala}  \&
  {Bottou}}{{Arjovsky} et~al.}{2017}]{wgan}
{Arjovsky} M.,  {Chintala} S.,   {Bottou} L.,  2017, arXiv e-prints, \href
  {https://ui.adsabs.harvard.edu/abs/2017arXiv170107875A} {p. arXiv:1701.07875}

\bibitem[\protect\citeauthoryear{Bishop}{Bishop}{1994}]{MDN}
Bishop C.,  1994, Workingpaper, Mixture density networks.
Aston University

\bibitem[\protect\citeauthoryear{{Blazek}, {MacCrann}, {Troxel}  \&
  {Fang}}{{Blazek} et~al.}{2019}]{blazek-tatt}
{Blazek} J.~A.,  {MacCrann} N.,  {Troxel} M.~A.,   {Fang} X.,  2019, \mn@doi
  [\prd] {10.1103/PhysRevD.100.103506}, \href
  {https://ui.adsabs.harvard.edu/abs/2019PhRvD.100j3506B} {100, 103506}

\bibitem[\protect\citeauthoryear{{Bridle} \& {King}}{{Bridle} \&
  {King}}{2007}]{bridle-king}
{Bridle} S.,  {King} L.,  2007, \mn@doi [New Journal of Physics]
  {10.1088/1367-2630/9/12/444}, \href
  {https://ui.adsabs.harvard.edu/abs/2007NJPh....9..444B} {9, 444}

\bibitem[\protect\citeauthoryear{{Catelan}, {Kamionkowski}  \&
  {Blandford}}{{Catelan} et~al.}{2001}]{catelan}
{Catelan} P.,  {Kamionkowski} M.,   {Blandford} R.~D.,  2001, \mn@doi [\mnras]
  {10.1046/j.1365-8711.2001.04105.x}, \href
  {https://ui.adsabs.harvard.edu/abs/2001MNRAS.320L...7C} {320, L7}

\bibitem[\protect\citeauthoryear{{Chisari} et~al.,}{{Chisari}
  et~al.}{2015}]{chisari-horizon-ia}
{Chisari} N.,  et~al., 2015, \mn@doi [\mnras] {10.1093/mnras/stv2154}, \href
  {https://ui.adsabs.harvard.edu/abs/2015MNRAS.454.2736C} {454, 2736}

\bibitem[\protect\citeauthoryear{{Davis}, {Efstathiou}, {Frenk}  \&
  {White}}{{Davis} et~al.}{1985}]{fof}
{Davis} M.,  {Efstathiou} G.,  {Frenk} C.~S.,   {White} S.~D.~M.,  1985,
  \mn@doi [\apj] {10.1086/163168}, \href
  {https://ui.adsabs.harvard.edu/abs/1985ApJ...292..371D} {292, 371}

\bibitem[\protect\citeauthoryear{Defferrard, Bresson  \&
  Vandergheynst}{Defferrard et~al.}{2016}]{Defferrard2016}
Defferrard M.,  Bresson X.,   Vandergheynst P.,  2016, in Proceedings of the
  30th International Conference on Neural Information Processing Systems.
  NIPS'16.
Curran Associates Inc., Red Hook, NY, USA, p. 3844–3852

\bibitem[\protect\citeauthoryear{{Dubois}, {Peirani}, {Pichon}, {Devriendt},
  {Gavazzi}, {Welker}  \& {Volonteri}}{{Dubois} et~al.}{2016}]{dubois-horizon}
{Dubois} Y.,  {Peirani} S.,  {Pichon} C.,  {Devriendt} J.,  {Gavazzi} R.,
  {Welker} C.,   {Volonteri} M.,  2016, \mn@doi [\mnras]
  {10.1093/mnras/stw2265}, \href
  {https://ui.adsabs.harvard.edu/abs/2016MNRAS.463.3948D} {463, 3948}

\bibitem[\protect\citeauthoryear{{Fortuna}, {Hoekstra}, {Joachimi}, {Johnston},
  {Chisari}, {Georgiou}  \& {Mahony}}{{Fortuna} et~al.}{2021}]{fortuna}
{Fortuna} M.~C.,  {Hoekstra} H.,  {Joachimi} B.,  {Johnston} H.,  {Chisari}
  N.~E.,  {Georgiou} C.,   {Mahony} C.,  2021, \mn@doi [\mnras]
  {10.1093/mnras/staa3802}, \href
  {https://ui.adsabs.harvard.edu/abs/2021MNRAS.501.2983F} {501, 2983}

\bibitem[\protect\citeauthoryear{{Goodfellow}, {Pouget-Abadie}, {Mirza}, {Xu},
  {Warde-Farley}, {Ozair}, {Courville}  \& {Bengio}}{{Goodfellow}
  et~al.}{2014}]{Goodfellow-GAN}
{Goodfellow} I.~J.,  {Pouget-Abadie} J.,  {Mirza} M.,  {Xu} B.,  {Warde-Farley}
  D.,  {Ozair} S.,  {Courville} A.,   {Bengio} Y.,  2014, arXiv e-prints, \href
  {https://ui.adsabs.harvard.edu/abs/2014arXiv1406.2661G} {p. arXiv:1406.2661}

\bibitem[\protect\citeauthoryear{{Gulrajani}, {Ahmed}, {Arjovsky}, {Dumoulin}
  \& {Courville}}{{Gulrajani} et~al.}{2017}]{wgangp}
{Gulrajani} I.,  {Ahmed} F.,  {Arjovsky} M.,  {Dumoulin} V.,   {Courville} A.,
  2017, arXiv e-prints, \href
  {https://ui.adsabs.harvard.edu/abs/2017arXiv170400028G} {p. arXiv:1704.00028}

\bibitem[\protect\citeauthoryear{{Guo} et~al.,}{{Guo} et~al.}{2011}]{guo-SAM}
{Guo} Q.,  et~al., 2011, \mn@doi [\mnras] {10.1111/j.1365-2966.2010.18114.x},
  \href {https://ui.adsabs.harvard.edu/abs/2011MNRAS.413..101G} {413, 101}

\bibitem[\protect\citeauthoryear{{Hearin} et~al.,}{{Hearin}
  et~al.}{2017}]{halotools}
{Hearin} A.~P.,  et~al., 2017, \mn@doi [\aj] {10.3847/1538-3881/aa859f}, \href
  {https://ui.adsabs.harvard.edu/abs/2017AJ....154..190H} {154, 190}

\bibitem[\protect\citeauthoryear{{Heitmann} et~al.,}{{Heitmann}
  et~al.}{2019}]{outer-rim}
{Heitmann} K.,  et~al., 2019, \mn@doi [\apjs] {10.3847/1538-4365/ab4da1}, \href
  {https://ui.adsabs.harvard.edu/abs/2019ApJS..245...16H} {245, 16}

\bibitem[\protect\citeauthoryear{{Hirata} \& {Seljak}}{{Hirata} \&
  {Seljak}}{2004}]{hirata-seljak}
{Hirata} C.~M.,  {Seljak} U.,  2004, \mn@doi [\prd]
  {10.1103/PhysRevD.70.063526}, \href
  {https://ui.adsabs.harvard.edu/abs/2004PhRvD..70f3526H} {70, 063526}

\bibitem[\protect\citeauthoryear{Ho, Rau, Ntampaka, Farahi, Trac  \&
  P{\'{o}}czos}{Ho et~al.}{2019}]{ho-halo-mass}
Ho M.,  Rau M.~M.,  Ntampaka M.,  Farahi A.,  Trac H.,   P{\'{o}}czos B.,
  2019, \mn@doi [The Astrophysical Journal] {10.3847/1538-4357/ab4f82}, 887, 25

\bibitem[\protect\citeauthoryear{{Horie}, {Morita}, {Hishinuma}, {Ihara}  \&
  {Mitsume}}{{Horie} et~al.}{2020}]{iso-gcn}
{Horie} M.,  {Morita} N.,  {Hishinuma} T.,  {Ihara} Y.,   {Mitsume} N.,  2020,
  arXiv e-prints, \href {https://ui.adsabs.harvard.edu/abs/2020arXiv200506316H}
  {p. arXiv:2005.06316}

\bibitem[\protect\citeauthoryear{{Jagvaral}, {Campbell}, {Mandelbaum}  \&
  {Rau}}{{Jagvaral} et~al.}{2021}]{gal_decomp}
{Jagvaral} Y.,  {Campbell} D.,  {Mandelbaum} R.,   {Rau} M.~M.,  2021, arXiv
  e-prints, \href {https://ui.adsabs.harvard.edu/abs/2021arXiv210502237J} {p.
  arXiv:2105.02237}

\bibitem[\protect\citeauthoryear{{Jagvaral}, {Singh}  \&
  {Mandelbaum}}{{Jagvaral} et~al.}{2022}]{IA-bulge-disc}
{Jagvaral} Y.,  {Singh} S.,   {Mandelbaum} R.,  2022, arXiv e-prints, \href
  {https://ui.adsabs.harvard.edu/abs/2022arXiv220208849J} {p. arXiv:2202.08849}

\bibitem[\protect\citeauthoryear{{Joachimi}, {Semboloni}, {Bett}, {Hartlap},
  {Hilbert}, {Hoekstra}, {Schneider}  \& {Schrabback}}{{Joachimi}
  et~al.}{2013}]{joachimi-2013}
{Joachimi} B.,  {Semboloni} E.,  {Bett} P.~E.,  {Hartlap} J.,  {Hilbert} S.,
  {Hoekstra} H.,  {Schneider} P.,   {Schrabback} T.,  2013, \mn@doi [\mnras]
  {10.1093/mnras/stt172}, \href
  {https://ui.adsabs.harvard.edu/abs/2013MNRAS.431..477J} {431, 477}

\bibitem[\protect\citeauthoryear{{Khandai}, {Di Matteo}, {Croft}, {Wilkins},
  {Feng}, {Tucker}, {DeGraf}  \& {Liu}}{{Khandai} et~al.}{2015}]{Khandai-mb2}
{Khandai} N.,  {Di Matteo} T.,  {Croft} R.,  {Wilkins} S.,  {Feng} Y.,
  {Tucker} E.,  {DeGraf} C.,   {Liu} M.-S.,  2015, \mn@doi [\mnras]
  {10.1093/mnras/stv627}, \href
  {https://ui.adsabs.harvard.edu/abs/2015MNRAS.450.1349K} {450, 1349}

\bibitem[\protect\citeauthoryear{{Kiessling} et~al.,}{{Kiessling}
  et~al.}{2015}]{ia-review2}
{Kiessling} A.,  et~al., 2015, \mn@doi [\ssr] {10.1007/s11214-015-0203-6},
  \href {https://ui.adsabs.harvard.edu/abs/2015SSRv..193...67K} {193, 67}

\bibitem[\protect\citeauthoryear{{Kilbinger}}{{Kilbinger}}{2015}]{Kilbinger}
{Kilbinger} M.,  2015, \mn@doi [Reports on Progress in Physics]
  {10.1088/0034-4885/78/8/086901}, \href
  {https://ui.adsabs.harvard.edu/abs/2015RPPh...78h6901K} {78, 086901}

\bibitem[\protect\citeauthoryear{{Kingma} \& {Ba}}{{Kingma} \&
  {Ba}}{2014}]{adam}
{Kingma} D.~P.,  {Ba} J.,  2014, arXiv e-prints, \href
  {https://ui.adsabs.harvard.edu/abs/2014arXiv1412.6980K} {p. arXiv:1412.6980}

\bibitem[\protect\citeauthoryear{{Kipf} \& {Welling}}{{Kipf} \&
  {Welling}}{2016}]{kipf-welling}
{Kipf} T.~N.,  {Welling} M.,  2016, arXiv e-prints, \href
  {https://ui.adsabs.harvard.edu/abs/2016arXiv160902907K} {p. arXiv:1609.02907}

\bibitem[\protect\citeauthoryear{{Kodi Ramanah}, {Wojtak}, {Ansari}, {Gall}  \&
  {Hjorth}}{{Kodi Ramanah} et~al.}{2020}]{kodi-a}
{Kodi Ramanah} D.,  {Wojtak} R.,  {Ansari} Z.,  {Gall} C.,   {Hjorth} J.,
  2020, \mn@doi [\mnras] {10.1093/mnras/staa2886}, \href
  {https://ui.adsabs.harvard.edu/abs/2020MNRAS.499.1985K} {499, 1985}

\bibitem[\protect\citeauthoryear{{Korytov} et~al.,}{{Korytov}
  et~al.}{2019}]{cosmodc2}
{Korytov} D.,  et~al., 2019, \mn@doi [\apjs] {10.3847/1538-4365/ab510c}, \href
  {https://ui.adsabs.harvard.edu/abs/2019ApJS..245...26K} {245, 26}

\bibitem[\protect\citeauthoryear{{Lee}, {Springel}, {Pen}  \& {Lemson}}{{Lee}
  et~al.}{2008}]{ed-ee}
{Lee} J.,  {Springel} V.,  {Pen} U.-L.,   {Lemson} G.,  2008, \mn@doi [\mnras]
  {10.1111/j.1365-2966.2008.13624.x}, \href
  {https://ui.adsabs.harvard.edu/abs/2008MNRAS.389.1266L} {389, 1266}

\bibitem[\protect\citeauthoryear{{Li}, {Ni}, {Croft}, {Di Matteo}, {Bird}  \&
  {Feng}}{{Li} et~al.}{2021}]{yin-li}
{Li} Y.,  {Ni} Y.,  {Croft} R. A.~C.,  {Di Matteo} T.,  {Bird} S.,   {Feng} Y.,
   2021, \mn@doi [Proceedings of the National Academy of Science]
  {10.1073/pnas.2022038118}, \href
  {https://ui.adsabs.harvard.edu/abs/2021PNAS..11822038L} {118, 2022038118}

\bibitem[\protect\citeauthoryear{{Mandelbaum} et~al.,}{{Mandelbaum}
  et~al.}{2011}]{mandelbaum-2011}
{Mandelbaum} R.,  et~al., 2011, \mn@doi [\mnras]
  {10.1111/j.1365-2966.2010.17485.x}, \href
  {https://ui.adsabs.harvard.edu/abs/2011MNRAS.410..844M} {410, 844}

\bibitem[\protect\citeauthoryear{Marinacci et~al.}{Marinacci
  et~al.}{2018}]{Marinacci2017illustristng}
Marinacci F.,  et~al., 2018, \mn@doi [Mon. Not. Roy. Astron. Soc.]
  {10.1093/mnras/sty2206}, 480, 5113

\bibitem[\protect\citeauthoryear{Mathieson \& Moscato}{Mathieson \&
  Moscato}{2019}]{proximity}
Mathieson L.,  Moscato P.,  2019, An Introduction to Proximity Graphs.
Springer International Publishing, Cham, pp 213--233,
  \mn@doi{10.1007/978-3-030-06222-4_4}, \url
  {https://doi.org/10.1007/978-3-030-06222-4_4}

\bibitem[\protect\citeauthoryear{{Miyato}, {Kataoka}, {Koyama}  \&
  {Yoshida}}{{Miyato} et~al.}{2018}]{sngan}
{Miyato} T.,  {Kataoka} T.,  {Koyama} M.,   {Yoshida} Y.,  2018, arXiv
  e-prints, \href {https://ui.adsabs.harvard.edu/abs/2018arXiv180205957M} {p.
  arXiv:1802.05957}

\bibitem[\protect\citeauthoryear{{Naiman} et~al.,}{{Naiman}
  et~al.}{2018}]{Naiman2018illustristng}
{Naiman} J.~P.,  et~al., 2018, \mn@doi [\mnras] {10.1093/mnras/sty618}, \href
  {https://ui.adsabs.harvard.edu/abs/2018MNRAS.477.1206N} {477, 1206}

\bibitem[\protect\citeauthoryear{Nelson et~al.}{Nelson
  et~al.}{2018}]{tng-bimodal}
Nelson D.,  et~al., 2018, \mn@doi [Mon. Not. Roy. Astron. Soc.]
  {10.1093/mnras/stx3040}, 475, 624

\bibitem[\protect\citeauthoryear{{Nelson} et~al.,}{{Nelson}
  et~al.}{2019a}]{nelson-tng-publicdata}
{Nelson} D.,  et~al., 2019a, \mn@doi [Computational Astrophysics and Cosmology]
  {10.1186/s40668-019-0028-x}, \href
  {https://ui.adsabs.harvard.edu/abs/2019ComAC...6....2N} {6, 2}

\bibitem[\protect\citeauthoryear{{Nelson} et~al.,}{{Nelson}
  et~al.}{2019b}]{tng-publicdata}
{Nelson} D.,  et~al., 2019b, \mn@doi [Computational Astrophysics and Cosmology]
  {10.1186/s40668-019-0028-x}, \href
  {https://ui.adsabs.harvard.edu/abs/2019ComAC...6....2N} {6, 2}

\bibitem[\protect\citeauthoryear{{Ntampaka} et~al.,}{{Ntampaka}
  et~al.}{2019}]{ntampaka-2019}
{Ntampaka} M.,  et~al., 2019, \baas, \href
  {https://ui.adsabs.harvard.edu/abs/2019BAAS...51c..14N} {51, 14}

\bibitem[\protect\citeauthoryear{{Pereira}, {Bryan}  \& {Gill}}{{Pereira}
  et~al.}{2008}]{radial-align}
{Pereira} M.~J.,  {Bryan} G.~L.,   {Gill} S. P.~D.,  2008, \mn@doi [\apj]
  {10.1086/523830}, \href
  {https://ui.adsabs.harvard.edu/abs/2008ApJ...672..825P} {672, 825}

\bibitem[\protect\citeauthoryear{{Pillepich} et~al.,}{{Pillepich}
  et~al.}{2018a}]{tng-methods}
{Pillepich} A.,  et~al., 2018a, \mn@doi [\mnras] {10.1093/mnras/stx2656}, \href
  {https://ui.adsabs.harvard.edu/abs/2018MNRAS.473.4077P} {473, 4077}

\bibitem[\protect\citeauthoryear{{Pillepich} et~al.,}{{Pillepich}
  et~al.}{2018b}]{pillepich2018illustristng}
{Pillepich} A.,  et~al., 2018b, \mn@doi [\mnras] {10.1093/mnras/stx3112}, \href
  {https://ui.adsabs.harvard.edu/abs/2018MNRAS.475..648P} {475, 648}

\bibitem[\protect\citeauthoryear{{Potter}, {Stadel}  \& {Teyssier}}{{Potter}
  et~al.}{2017}]{PKDGRAV3}
{Potter} D.,  {Stadel} J.,   {Teyssier} R.,  2017, \mn@doi [Computational
  Astrophysics and Cosmology] {10.1186/s40668-017-0021-1}, \href
  {https://ui.adsabs.harvard.edu/abs/2017ComAC...4....2P} {4, 2}

\bibitem[\protect\citeauthoryear{{Samuroff}, {Mandelbaum}  \&
  {Blazek}}{{Samuroff} et~al.}{2020}]{samuroff-2020}
{Samuroff} S.,  {Mandelbaum} R.,   {Blazek} J.,  2020, arXiv e-prints, \href
  {https://ui.adsabs.harvard.edu/abs/2020arXiv200910735S} {p. arXiv:2009.10735}

\bibitem[\protect\citeauthoryear{Satorras, Hoogeboom  \& Welling}{Satorras
  et~al.}{2021}]{egnn}
Satorras V.~G.,  Hoogeboom E.,   Welling M.,  2021, in Meila M.,  Zhang T.,
  eds,  Proceedings of Machine Learning Research Vol. 139, Proceedings of the
  38th International Conference on Machine Learning. PMLR, pp 9323--9332, \url
  {https://proceedings.mlr.press/v139/satorras21a.html}

\bibitem[\protect\citeauthoryear{{Schaye} et~al.,}{{Schaye}
  et~al.}{2015}]{Schaye-EAGLE}
{Schaye} J.,  et~al., 2015, \mn@doi [\mnras] {10.1093/mnras/stu2058}, \href
  {https://ui.adsabs.harvard.edu/abs/2015MNRAS.446..521S} {446, 521}

\bibitem[\protect\citeauthoryear{{Schneider} \& {Bridle}}{{Schneider} \&
  {Bridle}}{2010}]{schneider-bridle}
{Schneider} M.~D.,  {Bridle} S.,  2010, \mn@doi [\mnras]
  {10.1111/j.1365-2966.2009.15956.x}, \href
  {https://ui.adsabs.harvard.edu/abs/2010MNRAS.402.2127S} {402, 2127}

\bibitem[\protect\citeauthoryear{Singh, Mandelbaum, Seljak, Slosar  \&
  Vazquez~Gonzalez}{Singh et~al.}{2017}]{singh-covarience}
Singh S.,  Mandelbaum R.,  Seljak U.,  Slosar A.,   Vazquez~Gonzalez J.,  2017,
  \mn@doi [Monthly Notices of the Royal Astronomical Society]
  {10.1093/mnras/stx1828}, 471, 3827

\bibitem[\protect\citeauthoryear{{Somerville} \& {Dav{\'e}}}{{Somerville} \&
  {Dav{\'e}}}{2015}]{somerville-dave}
{Somerville} R.~S.,  {Dav{\'e}} R.,  2015, \mn@doi [\araa]
  {10.1146/annurev-astro-082812-140951}, \href
  {https://ui.adsabs.harvard.edu/abs/2015ARA&A..53...51S} {53, 51}

\bibitem[\protect\citeauthoryear{{Somerville}, {Hopkins}, {Cox}, {Robertson}
  \& {Hernquist}}{{Somerville} et~al.}{2008}]{somerville-SAM}
{Somerville} R.~S.,  {Hopkins} P.~F.,  {Cox} T.~J.,  {Robertson} B.~E.,
  {Hernquist} L.,  2008, \mn@doi [\mnras] {10.1111/j.1365-2966.2008.13805.x},
  \href {https://ui.adsabs.harvard.edu/abs/2008MNRAS.391..481S} {391, 481}

\bibitem[\protect\citeauthoryear{{Springel}}{{Springel}}{2010}]{arepo}
{Springel} V.,  2010, \mn@doi [\mnras] {10.1111/j.1365-2966.2009.15715.x},
  \href {https://ui.adsabs.harvard.edu/abs/2010MNRAS.401..791S} {401, 791}

\bibitem[\protect\citeauthoryear{{Springel}, {White}, {Tormen}  \&
  {Kauffmann}}{{Springel} et~al.}{2001}]{subfind}
{Springel} V.,  {White} S. D.~M.,  {Tormen} G.,   {Kauffmann} G.,  2001,
  \mn@doi [\mnras] {10.1046/j.1365-8711.2001.04912.x}, \href
  {https://ui.adsabs.harvard.edu/abs/2001MNRAS.328..726S} {328, 726}

\bibitem[\protect\citeauthoryear{Springel et~al.}{Springel
  et~al.}{2018}]{Springel2017illustristng}
Springel V.,  et~al., 2018, \mn@doi [Mon. Not. Roy. Astron. Soc.]
  {10.1093/mnras/stx3304}, 475, 676

\bibitem[\protect\citeauthoryear{{Tempel}, {Saar}, {Liivam{\"a}gi}, {Tamm},
  {Einasto}, {Einasto}  \& {M{\"u}ller}}{{Tempel} et~al.}{2011}]{env-sdss}
{Tempel} E.,  {Saar} E.,  {Liivam{\"a}gi} L.~J.,  {Tamm} A.,  {Einasto} J.,
  {Einasto} M.,   {M{\"u}ller} V.,  2011, \mn@doi [\aap]
  {10.1051/0004-6361/201016196}, \href
  {https://ui.adsabs.harvard.edu/abs/2011A&A...529A..53T} {529, A53}

\bibitem[\protect\citeauthoryear{{Tenneti}, {Mandelbaum}, {Di Matteo}, {Feng}
  \& {Khandai}}{{Tenneti} et~al.}{2014}]{tenneti-ia}
{Tenneti} A.,  {Mandelbaum} R.,  {Di Matteo} T.,  {Feng} Y.,   {Khandai} N.,
  2014, \mn@doi [\mnras] {10.1093/mnras/stu586}, \href
  {https://ui.adsabs.harvard.edu/abs/2014MNRAS.441..470T} {441, 470}

\bibitem[\protect\citeauthoryear{{Tenneti}, {Mandelbaum}  \& {Di
  Matteo}}{{Tenneti} et~al.}{2016}]{tenneti-disc-ellip}
{Tenneti} A.,  {Mandelbaum} R.,   {Di Matteo} T.,  2016, \mn@doi [\mnras]
  {10.1093/mnras/stw1823}, \href
  {https://ui.adsabs.harvard.edu/abs/2016MNRAS.462.2668T} {462, 2668}

\bibitem[\protect\citeauthoryear{Thomas, Smidt, Kearnes, Yang, Li, Kohlhoff  \&
  Riley}{Thomas et~al.}{2018}]{thomas2018tensor}
Thomas N.,  Smidt T.,  Kearnes S.,  Yang L.,  Li L.,  Kohlhoff K.,   Riley P.,
  2018, arXiv preprint arXiv:1802.08219

\bibitem[\protect\citeauthoryear{{Troxel} \& {Ishak}}{{Troxel} \&
  {Ishak}}{2015}]{troxel-ishak}
{Troxel} M.~A.,  {Ishak} M.,  2015, \mn@doi [\physrep]
  {10.1016/j.physrep.2014.11.001}, \href
  {https://ui.adsabs.harvard.edu/abs/2015PhR...558....1T} {558, 1}

\bibitem[\protect\citeauthoryear{{Velliscig} et~al.,}{{Velliscig}
  et~al.}{2015}]{eagle-ia}
{Velliscig} M.,  et~al., 2015, \mn@doi [\mnras] {10.1093/mnras/stv2198}, \href
  {https://ui.adsabs.harvard.edu/abs/2015MNRAS.454.3328V} {454, 3328}

\bibitem[\protect\citeauthoryear{Verma, Boyer  \& Verbeek}{Verma
  et~al.}{2017}]{Verma2017}
Verma N.,  Boyer E.,   Verbeek J.,  2017, ArXiv, abs/1706.05206

\bibitem[\protect\citeauthoryear{Villar, Hogg, Storey-Fisher, Yao  \&
  Blum-Smith}{Villar et~al.}{2021}]{villar2021scalars}
Villar S.,  Hogg D.~W.,  Storey-Fisher K.,  Yao W.,   Blum-Smith B.,  2021,
  Advances in Neural Information Processing Systems, 34, 28848

\bibitem[\protect\citeauthoryear{{Vogelsberger} et~al.,}{{Vogelsberger}
  et~al.}{2014}]{Vogelsberger-illustris}
{Vogelsberger} M.,  et~al., 2014, \mn@doi [\mnras] {10.1093/mnras/stu1536},
  \href {https://ui.adsabs.harvard.edu/abs/2014MNRAS.444.1518V} {444, 1518}

\bibitem[\protect\citeauthoryear{{Vogelsberger}, {Marinacci}, {Torrey}  \&
  {Puchwein}}{{Vogelsberger} et~al.}{2020}]{Vogelsberger-review}
{Vogelsberger} M.,  {Marinacci} F.,  {Torrey} P.,   {Puchwein} E.,  2020,
  \mn@doi [Nature Reviews Physics] {10.1038/s42254-019-0127-2}, \href
  {https://ui.adsabs.harvard.edu/abs/2020NatRP...2...42V} {2, 42}

\bibitem[\protect\citeauthoryear{{Wechsler} \& {Tinker}}{{Wechsler} \&
  {Tinker}}{2018}]{wechsler-tinker}
{Wechsler} R.~H.,  {Tinker} J.~L.,  2018, \mn@doi [\araa]
  {10.1146/annurev-astro-081817-051756}, \href
  {https://ui.adsabs.harvard.edu/abs/2018ARA&A..56..435W} {56, 435}

\bibitem[\protect\citeauthoryear{{Zhou} et~al.,}{{Zhou}
  et~al.}{2018}]{gnn-review}
{Zhou} J.,  et~al., 2018, arXiv e-prints, \href
  {https://ui.adsabs.harvard.edu/abs/2018arXiv181208434Z} {p. arXiv:1812.08434}

\makeatother
\end{thebibliography}

\appendix

 \section{Spectral Graph Convolutions}
\label{appendix:graph_appendix}

As an efficient alternative 
to a  full spectral graph  convolution, \cite{Defferrard2016} proposed to use parametric polynomial filters $g_\theta$, of the form:
\begin{equation}
	g_\theta(\mathbf{L}) = \sum_{k=0}^{K-1} \theta_k \mathbf{L}^k.
\end{equation}
where $\theta \in  \mathbb{R}^{K}$ is a vector of polynomial coefficients with $K$ specifying the polynomial order. 

Defining a filter in  terms of a polynomial function of the graph Laplacian has the advantage that the filter takes the same simple expression in Fourier space: 
\begin{equation}
	g_\theta(\mathbf{L}) =  \sum_{k=0}^{K-1} \theta_k (\mathbf{U} \mathbf{\Lambda}  \mathbf{U}^t)^k  = \sum_{k=0}^{K-1} \theta_k \mathbf{U} \mathbf{\Lambda}^k \mathbf{U}^t =  \mathbf{U}  g_\theta(\mathbf{\Lambda)} \mathbf{U}^t 
\end{equation}
A graph convolution with such a parametric filter can be defined as $g_\theta \star f = g_\theta(\mathbf{L}) f = \mathbf{U}  g_\theta(\mathbf{\Lambda)} \mathbf{U}^t f $.

Given this formalism, we find it useful to introduce Chebyshev decompositions, which for a function $f$ are given by 
\begin{equation}
	f(x) = \sum\limits_{k=0}^{\infty} b_{k} T_k(x).
\end{equation}
Here $T_k$ is the Chebyshev polynomial of order $k$. An efficient implementation of these graph convolutions can therefore be achieved by the Chebyshev approximation. Chebyshev polynomials form an orthogonal basis of $L^2([-1,1], dy/\sqrt{1 - y^2})$ and can be computed recursively as $T_k(x) = 2x T_{k-1}(x) - T_{k-2}(x)$ with $T_0 =1$, $T_1 = x$.  One can therefore define a spectral graph convolution in terms of a finite order Chebyshev polynomial:
\begin{equation}
	g_\theta = \sum_{k=0}^K \theta_k T_k(\mathbf{L}).
\end{equation}

Limiting this expansion to $K=1$ \citep{kipf-welling} yields the following expression:
\begin{equation}
	g_\theta \simeq \mathbf{\theta}_0 +  2 \mathbf{\theta}_1 \mathbf{L}
	\label{eq:approx_conv}
\end{equation}
Contrary to the full expression of the graph convolution by filter $g_\theta$, we see that this first order approximation no longer requires
computing the graph Fourier Transform and can be computed by a single application of the graph adjacency matrix.

\section{Testing the robustness of the fits}
\label{appendix:overfit}

\begin{figure} 
\includegraphics [width=3.3in]{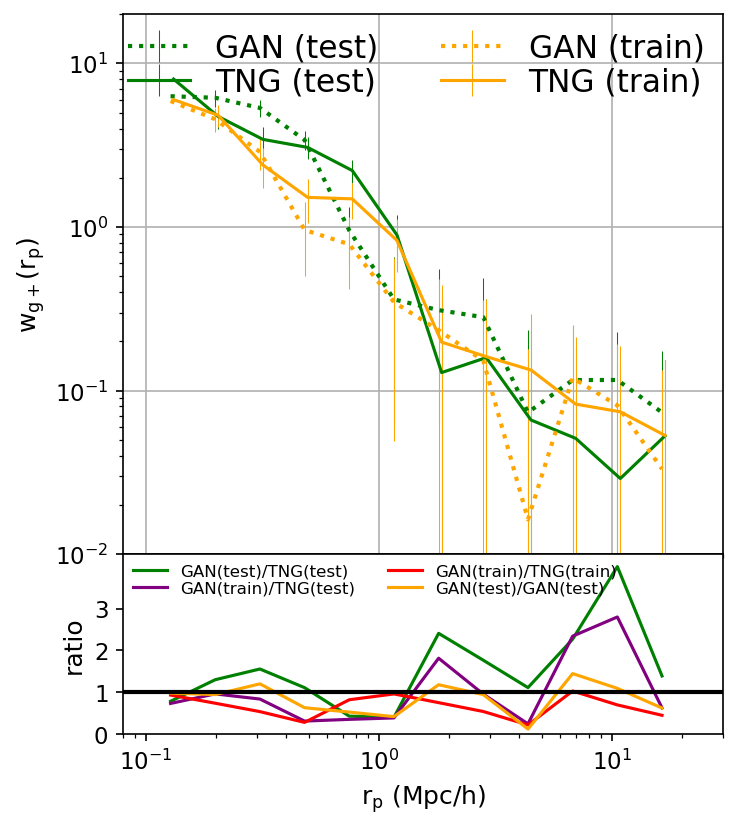}
 \caption{Projected two-point correlation functions $w_{g+}$ of galaxy positions and the projected 2D ellipticities of all galaxies, split into roughly equal-sized training and testing samples while preserving group membership. The top panel shows $w_{g+}$ measured using data from the TNG simulation in yellow and the data generated by the GAN in dotted green, while the bottom   panel shows the ratios among the curves as indicated by the label. All four curves are in good quantitative agreement, suggesting that the GAN is not significantly overfitting.  }\label{wgp_all_train_test}
 \end{figure}

In this appendix we investigate the robustness of our deep generative model. Due to scarcity of  cosmological  hydrodynamical simulations and the relatively limited volume available in TNG100, 
we used the whole sample to train and test.  Therefore, in this section we test whether the model has been overfit by splitting our sample roughly 50/50, while preserving group membership of subhalos and galaxies. In Fig.~\ref{wgp_all_train_test}, we present the model performance on both train and test samples.   

Additionally, we measured the \textit{Pearson-r} correlation coefficient between the generated  and the measured complex 2D ellipticities. We generated 100 different samples, each with a different random seed, measured the desired \textit{Pearson-r} coefficient and averaged the results. These were $ \langle \rho (\text{GAN} \, e_1, \text{TNG} \,  e_1)  \rangle = 0.09 $  and $\langle \rho (\text{GAN} \, e_2, \text{TNG} \, e_2) \rangle  = 0.06$.  The weak correlations suggest that the model did not just simply ``memorize'' the ellipticities.

  \section{Aleatoric (stochastic) uncertainty}
\label{appendix:aleatoric} 
\begin{figure} 
\includegraphics [width=3.3in]{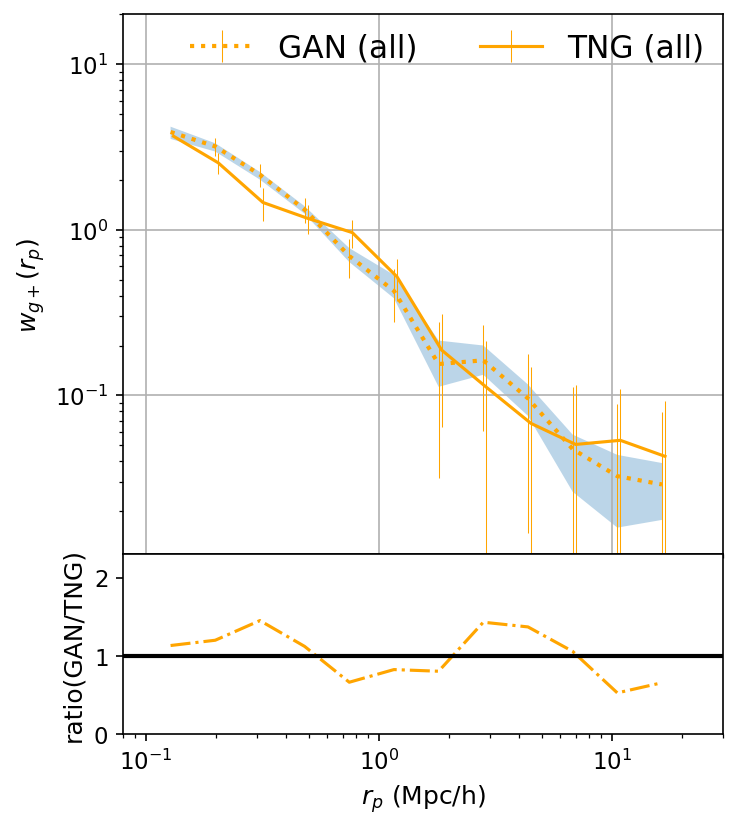}
 \caption{Projected two-point correlation function $w_{g+}$ of galaxy positions and the projected 2D ellipticities of all galaxies. The top panel shows  $w_{g+}$ measured using data from the TNG simulation in yellow and the data generated by the GAN in dotted green, while the bottom   panel shows the ratios among the curves as indicated by the label. In the blue shaded region we show the 1$\sigma$ scatter of $w_{g+}$  obtained using 50 different random seeds, as a measure of the stochastic uncertainty in the model.}\label{wgp_all_band}
 \end{figure}

 In this appendix we explore the uncertainty of the measured intrinsic alignment correlation functions due to the stochastic nature of the model implementation.\footnote{Note that due to the very complex nature of GANs and neural networks in general, it is usually very difficult to quantify model uncertainties, usually known as \textit{epistemic uncertainty}. Here, we do not attempt to model this type of uncertainty.}
 In Fig.~\ref{wgp_all_band}, we show the same quantity as in the first panel of Fig.~\ref{wgp_panels}, as well as its 1$\sigma$ scatter obtained using 50 different random seeds from the GAN. As is evident from the plot, this scatter is substantially smaller than the analytic errors that quantity the combination of shape noise and cosmic variance. Thus, we do not propagate and show this  random variance for every $w_{g+}$ curve. 
 \\
\bsp	
\label{lastpage}
\end{document}